\documentclass[11pt,onecolumn]{article}
\usepackage{setspace,dcolumn}
\usepackage{subfig}
\usepackage{hyperref}
\usepackage{cite}
\usepackage{graphicx,psfrag,epsfig,graphbox}
\usepackage[font=small,format=plain,labelfont=bf,justification=raggedright,singlelinecheck=false]{caption}
\usepackage{amsmath,amsfonts,amssymb,amsthm,bm,upgreek}
\allowdisplaybreaks[2]
\usepackage{geometry}
\usepackage[mathscr]{eucal}
\usepackage{esvect}
\usepackage{booktabs,threeparttable}
\usepackage{indentfirst}
\usepackage[normalem]{ulem}
\usepackage{lscape}
\usepackage{physics}
\usepackage{subfiles}
\usepackage{multirow}
\usepackage{tikz}
\usepackage{tikz-cd}

\usepackage{extarrows}

\usepackage{color}
\definecolor{darkgreen}{RGB}{40,150,60}
\definecolor{violet}{RGB}{140,50,230}
\definecolor{orange}{RGB}{230,150,0}

\newcommand{\wave}[1]{\widetilde{#1}}

\newcommand{\col}[2]{\begin{pmatrix}{#1}\\{#2}\end{pmatrix}}

\hypersetup{colorlinks=true,linkcolor=blue,anchorcolor=blue,citecolor=blue,urlcolor=blue}
\numberwithin{equation}{section}

\title{On self-dual Carrollian  conformal nonlinear electrodynamics}
\author{Bin Chen$^{1,2,3}$, Jue Hou$^{4}$, Haowei Sun$^2$}

\begin{document}
\maketitle
\begin{center}
	{\it
 $^1$ Department of Physics, School of Physical Science and Technology, Ningbo University, Ningbo, Zhejiang 315211, China\\\vspace{4mm}

		$^{2}$School of Physics, Peking University, No.5 Yiheyuan Rd, Beijing 100871, P.~R.~China\\
		\vspace{2mm}
		
		$^{3}$Center for High Energy Physics, Peking University, No.5 Yiheyuan Rd, Beijing 100871, P.~R.~China\\
		$^{4}$School of physics \& Shing-Tung Yau Center, Southeast University, Nanjing  211189,\\  P.~R.~China\\
	}
	\vspace{10mm}
\end{center}

\begin{abstract}
    \vspace{5mm}
    \begin{spacing}{1.5}
       In this work, we study the duality symmetry group of Carrollian (nonlinear) electrodynamics and propose a family of Carrollian ModMax theories, which are invariant under Carrollian  $\text{SO}(2)$ electromagnetic (EM) duality transformations and conformal transformation. We define the Carrollian $\text{SO}(2)$ EM transformations, with the help of Hodge duality in Carrollian geometry, then we rederive the  Gaillard-Zumino consistency condition for EM duality of Carrollian nonlinear electrodynamics. Together with the traceless condition for the energy-momentum tensor, we are able to determine the Lagrangian of the Carrollian ModMax theories among pure electrodynamics. We furthermore study their behaviors under the $\sqrt{T\bar{T}}$ deformation flow, and show that these theories deform to each other and may reach two endpoints under the flow, with one of the endpoint being the Carrollian Maxwell theory. As a byproduct, we construct a family of two-dimensional Carrollian ModMax-like multiple scalar theories, which are closed under the $\sqrt{T\bar{T}}$ flow and may flow to a BMS free multi-scalar model.
    \end{spacing}
\end{abstract}
\newpage

\setcounter{tocdepth}{2}
\tableofcontents

\newpage

\section{Introduction}\label{sec:Introduction}

    Electromagnetic (EM) duality is one of the most important features of Maxwell electrodynamics. {It reveals} the invariance of Maxwell’s equations under the transformation between electric fields and magnetic fields. The idea of EM duality was further developed to the Montonen–Olive duality \cite{Goddard:1976qe} and the S-duality in field theory and string theory\cite{Sen:1994fa}, which relates a strongly coupled theory to a weakly coupled one.  Although generally a duality relates two different theories, Maxwell theory in vacuum is said to be self-dual as the EM duality maps it to itself.  Moreover, Larmor \cite{larmor1925collected} and Rainich \cite{rainich1925electrodynamics} {noted that} the EM discrete duality transformation of Maxwell equations can be generalized to an one-parameter $\text{U}(1)$ rotation $(\mathbf{E}+i\mathbf{B})\to e^{-i\theta} (\mathbf{E}+i\mathbf{B})$. Such kind of generalized duality symmetry is also shared by many other nonlinear electrodynamics (NEDs) \cite{Gibbons:1995cv}, among which one notable example is the famous Born-Infeld theory. Most generally, as studied by Gaillard and Zumino in \cite{Gaillard:1981rj,Zumino:1981pt,Gaillard:1997zr,Gaillard:1997rt}, the duality symmetry group $\text{U}(1)$ (or $\text{SO}(2)$) can be extended to other subgroups of the general linear group $\text{GL}(2,\mathbb{R})$ in nonlinear electrodynamics interacting with matters, or subgroups of $\text{GL}(2n,\mathbb{R})$ in interacting theories of $n$ gauge fields. Among other cases, duality symmetries are also present in supersymmetric field theories\cite{Kuzenko:2000tg,Kuzenko:2000uh} and supergravity theories \cite{Ferrara:1976iq,Cremmer:1977tt,Cremmer:1978ds}.

    Apart from the duality rotation symmetry, the Maxwell theory is also conformal invariant. The existence of duality rotation invariant NEDs inspires naturally the question: is there any NED that preserves all the symmetries of the Maxwell theory? Historically, people gave a partially negative answer to this question and proved that the only possible Lorentz conformal invariant equations from the Lagrangians is the Maxwell electrodynamics, if the Lagrangian is a real analytic function of electromagnetic fields \cite{Bialynicki-Birula:1984daz,Bandos:2020jsw}. Recently, it was found that the Modified Maxwell (ModMax) theory \cite{Bandos:2020jsw,Kosyakov:2020wxv} presented a counter-example, in which case the Lagrangian and the Hamiltonian are not analytic functions. The ModMax theory is an one-parameter family of maximally symmetric nonlinear extensions of the Maxwell theory, in the sense that it is the unique Lorentz invariant deformation of the Maxwell theory which is both conformal invariant and $\text{SO}(2)$ duality rotation invariant, if the theory is defined {only in terms of the electromagnetic field strength.} The Lagrangian of ModMax is of the form
    \begin{equation}\label{eq:ModMaxLag-intro}
        \mathcal{L}_\gamma (\mathcal{S},\mathcal{P}) = -\frac{1}{2} \cosh{\gamma} \mathcal{S} + \frac{1}{2} \sinh{\gamma} \sqrt{\mathcal{S}^2+\mathcal{P}^2}; \qquad \mathcal{S}\equiv \frac{1}{2} F_{\mu\nu}F^{\mu\nu},\mathcal{P}\equiv \frac{1}{2} F_{\mu\nu}(*F)^{\mu\nu},
    \end{equation} 
    where $\gamma $ is a real parameter. 
    The free Maxwell electrodynamics can be viewed as the $\gamma=0$ ModMax, and at small $\gamma$ ModMax can be considered as an alternative to Maxwell theory, with some physical distinguishable implications such as vacuum birefringence \cite{Bandos:2020jsw}. Interestingly, the ModMax and its $2D$ cousin can be obtained from a $T\bar{T}$-like deformations of Maxwell electrodynamics \cite{Babaei-Aghbolagh:2022uij,Ferko:2022iru,Conti:2022egv,Ferko:2022cix,Babaei-Aghbolagh:2022leo,Morone:2024ffm}. The supersymmetric extension of ModMax was considered in \cite{Bandos:2021rqy,Kuzenko:2021cvx} and the higher-form extension was discussed in \cite{Bandos:2020hgy}. A manifestly duality symmetric democratic reformulation of the ModMax theory was obtained in \cite{Avetisyan:2021heg,Avetisyan:2022zza,Evnin:2023ypu}. The duality symmetry of ModMax extends to $\text{SL}(2,\mathbb{R})$ once coupled to an axion-dilaton field \cite{Babaei-Aghbolagh:2022itg}. The higher-derivative deformations of the ModMax theory is discussed in \cite{Kuzenko:2024zra}. Other recent works about ModMax include \cite{Shi:2024nmx,EslamPanah:2024tex,Karshiboev:2024xxx,Russo:2024llm,Ferko:2023iha,Neves:2022jqq}.

    It is notable that in most of the discussions on EM duality symmetries, these theories are supposed to be Lorentz invariant. However, the Lorentz invariance is not the only option. Actually, before the discovery of ModMax, it has been noticed in \cite{Bunster_2013} that there is a deep link between duality rotation invariance and the Poincar\'e symmetry, and  remarkably the zero signature contraction of the Poincar\'e group called Carrollian group is also allowed to have duality invariant electrodynamics.\footnote{Although in the original paper\cite{Bunster_2013} only the the Poincar\'e group and the Carrollian group {were considered, one may study the non-relativistic Galilean symmetry as well. In principle, there is no obstruction to obtain a Galilean version of ModMax by using our method, but the resulting theory has no interesting dynamics,} and we do not pay much attention to it in the main text. More comments will be given in the discussion section.} 
    The Carrollian symmetry group was first found by L\'evy-Leblond in 1965\cite{Levy-Leblond:1965} (and independently by Sen Gupta \cite{SenGupta:1966qer}) by studying the ultra-relativistic ($c \to 0$) contraction of the Poincar\'e group. The Carrollian physics  appear in various fields and has attracted  much attention especially in  the past decade. Some of the most recent researches include \cite{Aggarwal:2024gfb,Stieberger:2024shv,Bagchi:2024unl,Liu:2024nkc,He:2024yzx,Tadros:2024qlo,Banerjee:2024jub,Ecker:2023uwm,Mason:2023mti,Bergshoeff:2023vfd,Ciambelli:2023xqk,Kasikci:2023zdn,Marsot:2023qlc,deBoer:2023fnj,Nguyen:2023miw,Chen:2021xkw,Chen:2023pqf,Chen:2023esw,Chen:2023naw}. As far as we know,  except \cite{Bunster_2013} all other discussions about Carrollian NEDs \cite{Mehra:2024zqv,Ecker:2024czx} or Carrollian electrodynamics  did not touch the duality symmetry of the theory. It is demanding  to investigate the duality symmetry group of Carrollian (nonlinear) electrodynamics and the existence of a Carrollian version of ModMax.
    
    In this work, we try to study the  EM duality in $4D$ Carrollian electromagnetic theories. As first pointed out in \cite{Duval:2014uoa}, the Carrollian Maxwell theories are no longer invariant under the ultra-relativistic limit of the EM transformations. This is essentially because that the Hodge-$\ast$ operator is no longer well-defined for degenerate metrics. In order to discuss the EM duality in Carrollian geometry, we have to define Carrollian Hodge duality appropriately. We will show that the electric-type Carrollian Maxwell theory is self-dual under a modified discrete EM duality\footnote{One should distinguish this from the intertwining map \eqref{eq:CEDinterduality} first observed in \cite{Duval:2014uoa}.} and has a duality symmetry group isomorphic to $\mathbb{R}_{>0} \times \mathbb{R}$, but it is not invariant under a continuous  $\text{SO}(2)$  EM duality. By studying the Gaillard-Zumino condition for Carrollian nonlinear electrodynamics, we find a family of Carrollian ModMax theories, which is Carrollian  $\text{SO}(2)$  EM duality invariant as well as  conformal invariant. The Lagrangian of Carrollian ModMax theories is 
    \begin{equation}\label{eq:CarrModMax-intro}
        \mathcal{L}_{\gamma}(\mathcal{S},\mathcal{P}) = \frac{1}{4} \left(e^{\gamma}\mathcal{S} \mp e^{-\gamma}\frac{\mathcal{P}^2}{\mathcal{S}}\right); \qquad \mathcal{S} \equiv \frac{1}{2} F_{\mu\nu}F^{\mu\nu}=  m^{\mu\rho}\gamma^{\nu\sigma}F_{\mu\nu}F_{\rho\sigma}, \hspace{2ex}\mathcal{P} \equiv \frac{1}{2} F_{\mu\nu}\wave{F}^{\mu\nu}.
    \end{equation} It is remarkable that the Carrollian  $\text{SO}(2)$  EM duality is not the ultra-relativisitc limit of the usual $\text{SO}(2)$  EM duality. 

    Furthermore, we investigate the flow of the $4D$ Carrollian ModMax theories under the $\sqrt{T\bar{T}}$ deformation. In the relativistic case, the ModMax theories can be generated from the Maxwell theory by the $\sqrt{T\bar{T}}$ flow and the flow is totally invertible. However, we show a different picture in the Carrollian case: under $\sqrt{T\bar{T}}$ deformation, the Carrollian ModMax theories are closed in the sense that they are related to each other by the flow, and they can flow to  two endpoints, one of them being the electric-type Carrollian Maxwell theory, and the theories at the endpoints are the fixed points of the flow but have different duality symmetries. Conversely, the Carrollian ModMax theories cannot be generated from the Carrollian Maxwell theory, and the $\sqrt{T\bar{T}}$ deformation becomes non-invertible at the endpoints.

    As a byproduct of our study, we construct 2D Carrollian ModMax-like $N$-scalar theories and discuss their behaviors under the $\sqrt{T\bar{T}}$ flow. In the $N=2$ case, the Lagrangian of the theories can be obtained by dimensional reduction from $4D$ Carrollian ModMax theories. We generalize the Lagrangian to $N$-scalar and show that they are closed under $\sqrt{T\bar{T}}$ deformation. Similar to the Carrollian ModMax theories, these scalar theories flow to two fixed endpoints, with one of them being the BMS free $N$-scalar theory, and the flow at the endpoints is not invertible.  
    
    The remaining parts of this paper are organized as follows. In Section \ref{sec:CarrollianSymmety}, we briefly review the basics of Carrollian (conformal) symmetry and Carrollian geometry. In Section \ref{sec:HodgeStar}, we bring the ideas from \cite{Fecko:2022shq} to define Carrollian Levi-Civita tensors and the Carrollian Hodge-$*$ operator, and use it to discuss the discrete Hodge duality in the Carrollian Maxwell theory. In Section \ref{sec:GeneralizedDuality}, we generalize the discussion of continuous generalized duality symmetry to Carrollian NEDs. We find out the maximal duality group of the electric-type Carrollian Maxwell theory, and solve a one-parameter family of Carrollian NEDs - the Carrollian ModMax theories, which are both conformal and $\text{SO}(2)$ duality invariant. We also find the electric-type Carrollian Maxwell theory as one of the parameter space limits of the Carrollian ModMax theories. In Section \ref{sec:TTbar}, we investigate the behavior of $4D$ Carrollian ModMax theories under the $\sqrt{T\bar{T}}$ deformation flow, and generalize the study to a family of $2D$ modified BMS multiple-scalar model. The $4D$ Carrollian ModMax  provides another example to show that the duality-invariant NEDs may be generally related to $\sqrt{T\bar{T}}$-like flows, which {has been} proved for the relativistic case in \cite{Ferko:2023wyi}. We end with some discussions in Section \ref{Discussions}. We pack some technical details into an appendix.

    \paragraph{Conventions:} We use the Greek alphabets $\mu,\nu,\cdots$ as the spacetime indices, and use the lower-case Latin alphabets  $i,j, \cdots$ as spatial indices only. We use the upper-case alphabets $A, B,\cdots$ as internal indices in the tetrad formalism. Moreover, we use $a,b$ as the indices for $2D$ spacetime. When speaking of $D$-dimensional spacetime, we are talking about spacetime with a total dimension $D$. For Lorentzian spacetime, we select the almost positive signature $(-+++)$.\par
    
\section{Basics of Carrollian symmetry}\label{sec:CarrollianSymmety}
    We begin with a brief introduction to basic concepts about Carrollian symmetry and its conformal extension. Our focus is mainly on the $4d$ Carrollian electrodynamics, so we only discuss the global part of Carrollian symmetry and Carrollian conformal symmetry. However, it is worth noting that the global Carrollian conformal algebras in $d = 2, 3$ have extensions that are isomorphic to the infinite-dimensional BMS$_{d+1}$ algebra \cite{Duval:2014uva}. One may also refer to \cite{Chen:2021xkw} for further details about representations of higher dimensional Carrollian algebras, and \cite{Duval:2014uoa} for more details about the dual concepts of Newton-Cartan geometry and Carrollian geometry.\par
    
    \subsection{Carrollian symmetry and Carrollian conformal symmetry}\label{subsec:CCA}

    As the most famous non-trivial examples of kinematical algebras classified by Bacry and L\'evy-Leblond\cite{Bacry:1968zf}, both the Galilean and the Carrollian algebra emerge as the In\"on\"u--Wigner contractions of the Poincar\'e group, by taking the $c\to\infty$ (Galilean/non-relativistic) and the $c\to0$ (Carrollian/ultra-relativistic) limit respectively. After taking the Carrollian limit, the Lorentzian boosts become the Carrollian boosts:
    \begin{equation}\label{eq:GlobalBoost}
        \left\{
        \begin{aligned}
            t&\to t- \mathbf{b} \cdot \mathbf{x},\\
            \mathbf{x}&\to \mathbf{x}.\\
        \end{aligned}
        \right.
    \end{equation} At the same time, other parts of Poincar\'e group remain the same, so the Carrollian algebra is made up of the Carrollian boosts $B_i$, the spatial rotations $J_{ij}$ and the spacetime translations $P_\mu$.\par
    
    Conformal field theories (CFTs) are the theories with relativistic conformal symmetry and play a central role in modern theoretical physics. The Carrollian conformal symmetry naturally arises from the  $c\to0$ limit of relativistic conformal symmetry \cite{Bagchi:2016bcd}, and its symmetry algebra is called Carrollian conformal algebra (CCA), which is generated by the generators of the Carrollian algebra $\{P_\mu, J_{ij}, B_i\}$ plus the scaling generator $D$ and the Carrollian  special conformal transformations (SCTs) $K_\mu$. The generators in CCA correspond to the Carrollian conformal Killing vectors in Table \ref{tb:CCAAction}, and their commutation relations are:
    \begin{equation}\label{eq:CCACommutations}
        \begin{aligned}
            &[D,P_\mu]=P_\mu, ~~ [D,K_\mu]=-K_\mu, ~~ [D,B_i]=[D,J_{ij}]=0, \\
            &[J_{ij},G_k]=\delta_{ik}G_j-\delta_{jk}G_i, ~~ G\in\{P,K,B\}, ~~\\ &[J_{ij},P_0]=[J_{ij},K_0]=0,\\
            &[J_{ij},J_{kl}]=\delta_{ik}J_{jl}-\delta_{il}J_{jk}+\delta_{jl}J_{ik}-\delta_{jk}J_{il}, \\
            &[B_i,P_j]=\delta_{ij}P_0, ~~ [B_i,K_j]=\delta_{ij}K_0, ~~ [B_i,B_j]=[B_i,P_0]=[B_i,K_0]=0, \\
            &[K_0,P_0]=0, ~~ [K_0,P_i]=-2B_i, ~~ [K_i,P_0]=2B_i, ~~ [K_i,P_j]=2\delta_{ij}D+2J_{ij}.
        \end{aligned}
    \end{equation} Here $\mu=0,1,\dots,d-1$, and $i,j=1,\dots,d-1$.\par

    \begin{table}[ht]
        \def\arraystretch{1.6}
        \centering
        \caption{\centering The generators of the CCA and the Carrollian conformal Killing vector fields.}
        \label{tb:CCAAction}
        \begin{tabular}{clc}
            \hline
            Generators & \qquad Vector fields $\mathbf{v} = v^\mu \partial_{\mu}$\\
            \hline
            $D$ & \qquad$\mathbf{d} = t\partial_t+x^i\partial_i$ \\
            $P_\mu$ & \qquad$\mathbf{p}_\mu = \left(\partial_t ~, ~ \vec\partial\right)$ \\
            $K_\mu$ & \qquad$\mathbf{k}_\mu = \left(-\vec x^2\partial_t, 2\vec x x_\mu\partial^\mu-\vec x^2\vec\partial\right)$ \\
            $B_i$ & \qquad$\mathbf{b}_i = x^i\partial_t$ \\
            $J_{ij}$ & \qquad$\mathbf{j}_{ij} = x^i\partial_j-x^j\partial_i$ \\
            \hline
        \end{tabular}
    \end{table}\par

    Recently, different methods have been proposed to construct the field theories with Carrollian symmetry and Carrollian conformal symmetry. One can proceed by taking the Carrollian limit \cite{Bagchi:2016bcd,Banerjee:2020qjj,deBoer:2021jej,Henneaux:2021yzg}, or performing null reduction from Bargmann field theories \cite{Chen:2023pqf}, or using Galilean theories as seed theories \cite{Bergshoeff:2022qkx,deBoer:2023fnj}. Each of these methods allows us to find two different sectors \eqref{eq:CarrMaxwells} of Carrollian $\text{U}(1)$ gauge theory, which are referred to as the electric sector and the magnetic sector respectively,   and they are important examples of Carrollian conformal field theories.\par

    \subsection{Carrollian geometry}\label{subsec:Carrgeo}
    
    The relativistic field theory is defined on a Minkowski spacetime, and Einstein's general relativity is most conveniently defined in terms of (pseudo-)Riemann geometry.  It is natural to ask what kind of geometries underlying the Carrollian field theory and the Carrollian gravity are.  Over the years, the concept of Carrollian geometry  has been well established, both intrinsically \cite{Duval:2014uoa,Ciambelli:2019lap,Herfray:2021qmp,Figueroa-OFarrill:2022mcy} and by considering the Carrollian limit or small $c$ expansion of the Lorentzian geometry \cite{Bergshoeff:2017btm,Hansen:2021fxi}.\par
    
    The Carrollian geometry is described in a manner dual to the Newton-Cartan geometry which is used for non-relativistic gravity. A (weak) Carrollian structure is given by a triplet $(\mathcal{C},h_{\mu\nu},n^\mu)$ where $h_{\mu\nu}$ is a degenerate metric and the temporal vector $n^\mu$ is its kernel. Just like the Lorentz  group is the isometry group of the flat Minkowski spacetime, the Carrollian group is the isometry group of the \textbf{flat Carrollian structure} with
    \begin{equation}
        h_{\mu\nu}=\delta_{ij}dx^idx^j,\ n^\mu=\partial_0.
    \end{equation} Moreover, the Carrollian conformal group is the conformal group of the flat Carrollian structure, whose definition is
    \begin{equation}
        \varphi^* h = \Omega^2 h, \qquad \varphi_* n = \Omega^{-2/k}n.
    \end{equation} Here $\varphi$ is the spacetime transformation associated to a group element and $\Omega(x)$ is the conformal factor. Usually, the Carrollian conformal symmetry has $k=2$ to fix the scaling in the $n^\mu$ direction to be the same as the ones in other directions.\par

    We will denote $(n^\mu,h_{\mu\nu})$ as the Carrollian structure, and $(\tau_\mu,\gamma^{\mu\nu})$ for its inverse. The meaning of the inverse is that they satisfy the following relations:
    \begin{equation}
        \begin{aligned}
            &n^\mu h_{\mu\nu} = \tau_{\mu}\gamma^{\mu\nu}=0, \quad n^\mu\tau_{\mu}=1,\\
            &n^\mu\tau_{\nu} + \gamma^{\mu\rho}h_{\rho\nu} = \delta^{\mu}_{\ \nu}.\\
        \end{aligned}
    \end{equation}
     Since only the Carrollian structure is relevant to the physics, any Carrollian physical quantity needs to be invariant under a gauge symmetry generated by a coordinate dependent covector $b_\mu(x)$ called the local boost transformation, acting on the inverse structure $(\tau_\mu,\gamma^{\mu\nu})$ as:
    \begin{equation}\label{eq:LocalBoost}
        \tau_\mu \to \tau_\mu + b_\mu, \quad \gamma^{\mu\nu} \to \gamma^{\mu\nu} + 2 b^{(\mu}n^{\nu)} \qquad \text{where} \qquad n^\mu b_\mu=0, \quad b^\mu = \gamma^{\mu\nu}b_\nu.
    \end{equation}
    The reason is that the Carrollian structure $(n^\mu,h_{\mu\nu})$ does not have a unique inverse, and the physics should not depend on the choice of the inverse. If $(\tau_\mu,\gamma^{\mu\nu})$ is an inverse of $(n^\mu,h_{\mu\nu})$,  any $(\tau'_\mu,\gamma'^{\mu\nu})$ related to it by \eqref{eq:LocalBoost} is also an inverse, {but} the physics should be the same for both of them. One should distinguish the local boost gauge symmetry \eqref{eq:LocalBoost} from the spacetime Carrollian boost transformation \eqref{eq:GlobalBoost} which is global. Actually, the existence of this extra gauge symmetry is a significant feature of the non-Riemannian geometry which also includes the non-relativistic Newton-Cartan geometry, where the local boost transformation is called the Milne boost. One may refer to \cite{Hartong:2022lsy} for more details about the Newton-Cartan geometry.

    One can also describe  the Carrollian geometry in the tetrad formalism. In this formalism, the Carrollian structure and its inverse  can be  reorganized into the tetrad $\mathcal{E}^A_\mu = (\tau_\mu, e^i_{\mu})$ and its inverse, the cotetrad $\mathcal{E}^\mu_A =(n^\mu, e_i^{\mu})$,
    \begin{equation}
        \mathcal{E}^A_\nu\mathcal{E}_A^\mu = \delta^\mu_{\ \nu}, \quad \mathcal{E}^A_\mu\mathcal{E}_B^\mu = \delta^A_{\ B}.
    \end{equation} The local Carrollian symmetry acts on the tetrads by 
    \begin{equation}     
        \mathcal{E}^A_\mu \to C^A_{\ B}\mathcal{E}^B_\mu,\hspace{3ex}
            \mathcal{E}^\mu_A \to (C^{-1})^{\ B}_{A} \mathcal{E}^\mu_B,   
    \end{equation} 
    where 
    \begin{equation}
      C = \begin{pmatrix}
        1&\mathbf{b}^\intercal\\0&\mathbf{R}\in \text{SO}(d-1)\\
    \end{pmatrix} \in \text{Carr}_0(d)  
    \end{equation}
    is the $(d \cross d)$ representation matrix of an element in the { isometry} subgroup $\text{Carr}_0(d)$ of the $d$-dimensional Carrollian group, which satisfies $\det C = 1$. Consequently, the determinant of the tetrad matrix $\det(\mathcal{E}^A_\mu)$ is invariant under the local Carrollian symmetry. Last, the Carrollian metric is recovered by $h_{\mu\nu} = e^i_{\mu}e^i_{\nu}$. Different from that the local Lorentz symmetry acts trivially on the Lorentz metric, the local Carrollian boost symmetry acts non-trivially on the Carrollian structure. It can be easily checked that this transformation is consistent with the above local boost transformation \eqref{eq:LocalBoost}, which provides another origin of the local boost symmetry \eqref{eq:LocalBoost}. From now on, we speak of Carrollian invariance by requiring both the Carrollian spacetime transformation invariance and the local boost invariance.\par
    
\section{Non-Riemannian Hodge duality}\label{sec:HodgeStar}
    In four-dimensional ($4D$) relativistic Maxwell theory, the equations of motion are Maxwell’s equations
    \begin{equation}\label{eq:Maxwell}
        \begin{aligned}
        \left\{
        \begin{aligned}
            \nabla \cdot \mathbf{E} &= 0,  &\nabla \times \mathbf{B} - \frac{1}{c^2}\frac{\partial\mathbf E}{\partial t}&= 0,& \\
            \nabla \cdot \mathbf{B} &= 0,  &\nabla \times \mathbf{E} +\frac{\partial\mathbf B}{\partial t}&= 0.&
        \end{aligned}
        \right. \quad \text{(Maxwell)} 
        \end{aligned}
    \end{equation} 
    The well-known electromagnetic duality exchanges the electric and magnetic fields
    \begin{equation}\label{eq:MaxwellEMDual}
        \mathbf{E} \to c\mathbf{B}, \qquad c\mathbf{B} \to -\mathbf{E},
    \end{equation}
    and is a symmetry of Maxwell’s equations. 
    This duality map can be interpretated as the map between the covariant field strength 2-form $F_{\mu\nu}=2\partial_{[\mu}A_{\nu]}$ and its ``Hodge-$*$" dual: 
    \begin{equation}
        F_{\mu\nu} \to (*F)_{\mu\nu}, \qquad (*F)_{\mu\nu} \to -F_{\mu\nu}.
    \end{equation} \par
    
    The Carrollian cousins of Maxwell theory, commonly referred to as Carrollian electrodynamics or Carrollian Maxwell theories, have been studied extensively \cite{Duval:2014uoa, Chen:2023pqf,Bagchi:2016bcd,Banerjee:2020qjj,deBoer:2021jej,Bergshoeff:2022qkx,deBoer:2023fnj,Henneaux:2021yzg,Basu:2018dub}.  It turns out that there are two types of Carrollian Maxwell’s equations, the electric-type and the magnetic-type, 
    \begin{equation}\label{eq:CarrMaxwells}
        \begin{aligned}
        \left.\left\{
        \begin{aligned}
            \nabla \cdot \mathbf{E}_e &= 0, & \frac{\partial\mathbf{E}_e}{\partial t}&= 0,& \\
            \nabla \cdot \mathbf{B}_e &= 0, & \nabla \times \mathbf{E}_e +\frac{\partial\mathbf{B}_e}{\partial t}&= 0.&
        \end{aligned}
        \right.  \text{(elec.)} \quad \right| \left\{
        \begin{aligned}
            \nabla \cdot \mathbf{E}_m &= 0, & \nabla \times \mathbf{B}_m - \frac{\partial\mathbf{E}_m}{\partial t}&= 0,& \\
            \nabla \cdot \mathbf{B}_m &= 0, & \frac{\partial\mathbf{B}_m}{\partial t}&=0.&
        \end{aligned}
        \right.  \text{(mag.)}
        \end{aligned}
    \end{equation}
    These two sectors arise either as two different limits of Maxwell’s equations under suitable rescaling of $c$,
    \begin{equation}\label{eq:UR-limit}
        \begin{aligned}
            &\mathbf{E}_e &=& &\mathbf{E},& \qquad \mathbf{B}_e &=& &\mathbf{B}&,\\
            &\mathbf{E}_m &=& &c^{-1}\mathbf{E},& \qquad \mathbf{B}_m &=& &c\mathbf{B}&,\\
        \end{aligned}
    \end{equation}
    or as the equations of motions of the theories obtained by doing null reduction from two different Bargmann  $\text{U}(1)$  gauge theories\cite{Chen:2023pqf}.

    As first pointed out in \cite{Duval:2014uoa}, the Carrollian electrodynamics are no longer invariant under the ultra-relativistic limit \eqref{eq:UR-limit} of the (discrete) electric-magnetic duality transformation \eqref{eq:MaxwellEMDual}:
    \begin{equation}\label{eq:CEDinterduality}
        \mathbf{E}_e \to \mathbf{B}_m, \qquad \mathbf{B}_e \to -\mathbf{E}_m.
    \end{equation} 
    Instead, this transformation intertwines the electric-type equations with the magnetic-type ones in \eqref{eq:CarrMaxwells}.
    
    The breaking of electric-magnetic duality is essentially the consequence of the fact that the standard Hodge-$*$ operator is no longer well-defined for degenerate metrics. Nevertheless, one may generalize the definition of Hodge-$*$ operator in Riemann geometry to both Galilean and Carrollian spacetimes\cite{Fecko:2022shq}. We will review the definition of the Carrollian Hodge-$\ast$ operator, and moreover show that with the help of the generalized Carrollian Hodge-$*$ operator, even though the electric-type\footnote{For magnetic-type theory, we can still find a similar duality transformation which preserves the magnetic-type Carrollian Maxwell’s equations. However, it will be much more subtle to treat it in a fully Carrollian covariant manner compared to the electric-type theory. As can be seen in \cite{Chen:2023pqf}, the Lagrangian description of the magnetic-type theory is more complicated because  some additional auxiliary fields should be introduced. So in this sense, the magnetic-type theory is not a ``pure electrodynamics", whose Lagrangian $\mathcal{L}(F)$  depends only on $F_{\mu\nu}$. See more in the discussion part.} Carrollian electrodynamics is not strictly ``self-dual" under the duality transformation rule given by the modified Hodge-$*$ operator
    \begin{equation}\label{eq:CarrEMDual}
        F_{\mu\nu} \to (*F)_{\mu\nu}, \qquad (*F)_{\mu\nu} \to (*^2 F)_{\mu\nu}=0,
    \end{equation} 
    the theory can be invariant under a modified duality transformation 
    \begin{equation}\label{eq:CarrEMDual-new}
        F_{\mu\nu} \to F_{\mu\nu} + (*F)_{\mu\nu}, \qquad (*F)_{\mu\nu} \to (*F)_{\mu\nu}.
    \end{equation}\par

    \subsection{Non-Riemannian Levi-Civita tensors}

    We first review the classical theory of the Hodge-$*$ operator on a (pseudo-)Riemann manifold. The Hodge-$*$ operator is a linear map defined on the space of differential forms $\Omega^*(\mathcal{M})$ of a (pseudo-)Riemann manifold $\mathcal{M}$ which is endowed with a non-degenerate metric $g_{\mu\nu}$ and its inverse $g^{\mu\nu}$. Conventionally, we use $|g|$ for the determinant of $g$ and $\norm{g}$ for the absolute value of the determinant. On a $d$-dimensional manifold, the Hodge-$*$ operator maps $k$-forms to $(d-k)$-forms. Practically, the Levi-Civita tensor $\varepsilon_{\mu_1\cdots\mu_d}$ (or lifted to $\varepsilon^{\mu_1\cdots\mu_d}$ by $g^{\mu\nu}$) is used to explicitly write down the Hodge dual of a $k$-form $A_{\nu_1\cdots\nu_p}$: 
    \begin{equation}\label{eq:LorentzLCtensor}
        \begin{aligned}
            & \varepsilon_{\mu_1\cdots\mu_d} \equiv \sqrt{\norm{g}} \wave{\varepsilon}_{\mu_1\cdots\mu_d}, \quad \text{where} \quad \wave{\varepsilon}_{\mu_1\cdots\mu_d} = \wave{\varepsilon}^{\mu_1\cdots\mu_d} = \left\{ 
            \begin{aligned}
            +&1  &(\mu_1\cdots\mu_d)&\text{ is an even permutation}\\
            -&1  &(\mu_1\cdots\mu_d)&\text{ is an odd permutation}\\
            &0   &\text{otherwise}
            \end{aligned}
            \right. ,\\
            & \varepsilon^{\mu_1\cdots\mu_d} = \ g^{\mu_1\nu_1} \cdots g^{\mu_d\nu_d} \ \varepsilon_{\nu_1\cdots\nu_d}  = (-1)^s \sqrt{\norm{g}}^{-1}  \wave{\varepsilon}_{\mu_1\cdots\mu_d}, \qquad \text{$s$ is the signature of $g$}, \\
            & (*A)_{\mu_1\cdots\mu_{d-p}} \equiv \frac{1}{p!} \ g^{\rho_1\nu_1} \cdots g^{\rho_p\nu_p} \ \varepsilon_{\mu_1\cdots\mu_{d-p}\rho_1\cdots\rho_p}A_{\nu_1\cdots\nu_p}.
        \end{aligned}       
    \end{equation}\par
    
    Geometrically, the Levi-Civita tensor defines an orientation that is compatible with the (pseudo-)Riemann metric. Although a non-degenerate metric $g_{\mu\nu}$ is necessary for the standard definition of the Hodge dual of a differential form, the concept of Levi-Civita tensor can be generalized to the situations where the spacetime is only an oriented manifold, {not necessarily a (pseudo-)Riemann manifold. It could  } be the covariant \textit{orientation form} (also named \textit{volume form}) of the manifold compatible with other kinds of geometric data such as the Carrollian structure. \par

    Fortunately, there exist two  Carrollian covariant and locally boost invariant tensors in Carrollian spacetime \cite{Fecko:2022shq}, one being lower indexed and the other being upper indexed, which can be used as the Carrollian Levi-Civita tensors. Formally, one of these Carrollian Levi-Civita tensors for Carrollian structure $(n^\mu,h_{\mu\nu})$ is
    \begin{equation}\label{eq:CarrLCTensor-1}
        \varepsilon_{\mu_1\cdots\mu_d} \equiv \sqrt{\norm{g_\tau}} \wave{\varepsilon}_{\mu_1\cdots\mu_d} =\sqrt{(-1)^s|g_\tau|} \wave{\varepsilon}_{\mu_1\cdots\mu_d}, \qquad (g_\tau)_{\mu\nu} \equiv (-1)^s \tau_\mu\tau_\nu + h_{\mu\nu}.
    \end{equation} It is the same as the standard Levi-Civita tensor of a non-degenerate metric $g_\tau$. The metric $g_\tau$ depends on the selection of $(\tau_\mu,\gamma^{\mu\nu})$ so it is not locally boost invariant, but the determinant of $g_\tau$ is independent of this choice and is locally boost invariant. This can be most easily seen in the tetrad formalism  introduced in Section \ref{subsec:Carrgeo} because the determinant $|g_\tau|$ is the square of the tetrad determinant $\det(\mathcal{E}^A_{\mu})$. The signature $(-1)^s$ of the Carrollian manifold contributes only a sign in the determinant and does not have an exact meaning as in the Lorentz case, and can be chosen among $\pm 1$ arbitrarily. Though $(-1)^s=-1$ is more convenient to compare with the ultra-relativistic limit of Lorentz theory, in this paper we use $(-1)^s=1$ for simplicity.\par

    By considering the dual of a volume form, we have the following $n$-vector, which is also  Carrollian covariant and locally boost invariant,
    \begin{equation}\label{eq:CarrLCVector-2}
        \varepsilon^{\mu_1\cdots\mu_d} \equiv (-1)^s \sqrt{\norm{g_\gamma}}^{-1} \wave{\varepsilon}^{\mu_1\cdots\mu_d} , \qquad (g_\gamma)^{\mu\nu} \equiv (-1)^s n^\mu n^\nu + \gamma^{\mu\nu}.
    \end{equation} We also call it the Carrollian Levi-Civita tensor. The relation between this tensor and the volume form \eqref{eq:CarrLCTensor-1} is 
    \begin{equation}
        \varepsilon^{\mu_1\cdots\mu_d} = \ g_\gamma^{\mu_1\nu_1} \cdots g_\gamma^{\mu_d\nu_d} \ \varepsilon_{\nu_1\cdots\nu_d}, \qquad \varepsilon_{\mu_1\cdots\mu_d} = \ (g_\tau)_{\mu_1\nu_1} \cdots (g_\tau)_{\mu_d\nu_d} \ \varepsilon^{\nu_1\cdots\nu_d}.
    \end{equation}
    
    \subsection{Carrollian Hodge-$*$ and electromagnetic duality}
    With the help of the Carrollian Levi-Civita tensor, one may formally define the Carrollian Hodge-$*$ operator as
    \begin{equation}\label{eq:CarrHodgeStar}
        \begin{aligned}
            (*A)_{\mu_1\cdots\mu_{d-p}} &\equiv \frac{1}{p!} \ g_\gamma^{\rho_1\nu_1} \cdots g_\gamma^{\rho_p\nu_p} \ \varepsilon_{\mu_1\cdots\mu_{d-p}\rho_1\cdots\rho_p}A_{\nu_1\cdots\nu_p} \\
            &= \frac{1}{p!} \ (g_\tau)_{\mu_1\nu_1} \cdots (g_\tau)_{\mu_{d-p}\nu_{d-p}} \ \varepsilon^{\nu_1\cdots\nu_{d-p}\rho_1\cdots\rho_p}A_{\rho_1\cdots\rho_p}. 
        \end{aligned}
    \end{equation} However, this is not a good definition because $(g_\tau)_{\mu\nu},\ g_\gamma^{\mu\nu}$ enters so it depends on the choice of $(\tau_\mu,\gamma^{\mu\nu})$. To make the definition locally boost invariant, we can replace $(g_\tau)_{\mu\nu},\ g_\gamma^{\mu\nu}$ with the locally boost invariant quantities $h_{\mu\nu},\ m^{\mu\nu} \equiv (-1)^s n^\mu n^\nu$, and the two lines in \eqref{eq:CarrHodgeStar}  no longer equal. Due to the antisymmetric nature of the Levi-Civita tensor, the contraction of $\varepsilon_{\mu_1\cdots\mu_d}$ with more than two $m^{\mu\nu}$-s vanishes. Thus the more interesting choice is to use $\varepsilon^{\mu_1\cdots\mu_d}$ and $h_{\mu\nu}$ to define
    \begin{equation}\label{eq:CarrHodgeStar-new}
        (*A)_{\mu_1\cdots\mu_{d-p}} \equiv  \frac{1}{p!} \ h_{\mu_1\nu_1} \cdots h_{\mu_{d-p}\nu_{d-p}} \ \varepsilon^{\nu_1\cdots\nu_{d-p}\rho_1\cdots\rho_p}A_{\rho_1\cdots\rho_p}. 
    \end{equation} We will use this as our definition of the Carrollian Hodge-$*$ operator, {which is} one of the two Hodge stars defined in \cite{Fecko:2022shq}.\par

    Different from the (anti-)idempotent property $*^2=(-1)^s$ of standard Hodge-$*$ operator, the Carrollian Hodge-$*$ operator defined above is nilpotent, $*^2=0$. Then in electromagnetic duality if we take $F_{\mu\nu} \to (*F)_{\mu\nu}$, for self-consistence we are only allowed to choose $(*F)_{\mu\nu} \to 0$ instead of $(*F)_{\mu\nu} \to (-1)^s F_{\mu\nu}$. { Using $F_{\mu\nu} = E_e^i dx^i \wedge dt+ \frac{1}{2} \varepsilon_{ijk}B_e^idx^j\wedge dx^k$, the $4D$ electric-type Carrollian Maxwell’s equations in \eqref{eq:CarrMaxwells} can be rewritten in terms of} the Carrollian Hodge-$*$ and exterior derivative $d$ as
    \begin{equation}\label{eq:ElecHodgeCarrMaxwell}
        d*F = 0, \qquad dF=0.
    \end{equation} It is easy to check that the equations degenerate under the electromagnetic duality \eqref{eq:CarrEMDual} and are not strictly self-dual. However, upon adding an extra $F_{\mu\nu}$ into the transformation such that  $F_{\mu\nu} \to F_{\mu\nu}+(*F)_{\mu\nu}$, we have $(*F)_{\mu\nu} \to (*F)_{\mu\nu}$, as  shown in \eqref{eq:CarrEMDual-new}, the full electric-type Carrollian Maxwell’s equations are  self-dual under this transformation.\par

\section{Generalized duality symmetry of Carrollian nonlinear electrodynamics}\label{sec:GeneralizedDuality}
    In Section \ref{sec:HodgeStar} we discussed the discrete duality symmetry of Carrollian electrodynamics. For relativistic Maxwell theory, it is well-known that it is conformal invariant as well and the discrete duality symmetry can be generalized to a continuous $\text{SO}(2)$ EM duality symmetry:
    \begin{equation}\label{eq:MaxwellSO(2)Dual}
        \mathbf{E} \to \cos{\theta} \ \mathbf{E} + \sin{\theta} \ c\mathbf{B}, \qquad c\mathbf{B} \to \cos{\theta} \ c\mathbf{B} - \sin{\theta} \ \mathbf{E}.
    \end{equation} Actually, the Maxwell theory is not the unique theory with $\text{SO}(2)$ EM duality symmetry.  The duality rotation symmetry has been found in many other nonlinear electromagnetic theories (NEDs) such as the famous Born-Infeld theory. A few years ago\cite{Bandos:2020jsw, Kosyakov:2020wxv}, a one-parameter family of NEDs was discovered to share almost the same symmetries as the Maxwell theory. {This so-called ModMax (Modified Maxwell) theory has a Lagrangian}
    \begin{equation}\label{eq:ModMaxLag}
        \mathcal{L}_\gamma (\mathcal{S},\mathcal{P}) = -\frac{1}{2} \cosh{\gamma} \mathcal{S} + \frac{1}{2} \sinh{\gamma} \sqrt{\mathcal{S}^2+\mathcal{P}^2}; \qquad \mathcal{S}\equiv \frac{1}{2} F_{\mu\nu}F^{\mu\nu},~~\mathcal{P}\equiv \frac{1}{2} F_{\mu\nu}(*F)^{\mu\nu},
    \end{equation}
    where $\gamma \in \mathbb{R}$ is a real free parameter. Together with the Bialynicki-Birula theory, they are the only NED theories with both conformal symmetry and an $\text{SO}(2)$ EM duality symmetry. Later, the intriguing relation between the relativistic ModMax theories in various dimensions  and the $\sqrt{T\bar{T}}$ flow was exploited in \cite{Babaei-Aghbolagh:2022uij, Ferko:2022iru, Ferko:2022cix, Conti:2022egv, Babaei-Aghbolagh:2022leo}.\par

    In this section, we first derive the Carrollian Gaillard-Zumino condition for generalized duality symmetry of NEDs by following the original relativistic formalism in \cite{Gaillard:1981rj}. Then making use of this condition, we investigate the generalized duality symmetry of electric-type Carrollian Maxwell theory. Furthermore, though electric-type Carrollian Maxwell theory lacks $\text{SO}(2)$ EM duality symmetry, we discover a family of Carrollian conformal NED theories that are $\text{SO}(2)$ EM duality invariant. {This new family of theories will be referred to as  Carrollian ModMax theories. We show that the electric-type Carrollian Maxwell theory  can be obtained as the limiting theories of this new family.} We would like to point out that the Carrollian ModMax theories we get here are NOT the ultra-relativistic limit of the Lorentz ModMax theories. {Actually, taking the Carrollian limit directly, as the Galilean limit of ModMax taken in \cite{Banerjee:2022sza},  breaks  up the $\text{SO}(2)$ EM duality symmetry.} For a review of generalized duality symmetries and their applications, one may refer to \cite{Aschieri:2008zz}. In \cite{Sorokin:2021tge}, an overview of general properties of nonlinear electrodynamics, in particular, the ModMax electrodynamics, is given.\par

    \subsection{Gaillard-Zumino condition for Carrollian NEDs}
    
    It is illuminating to start from  the covariant formalism of NED \cite{Aschieri:2008zz}. In this formalism, one may introduce the concept of dual field strength\footnote{In some literature the definition here is replaced by $\wave{G}^{\mu\nu} \equiv 2\frac{\partial}{\partial F_{\mu\nu}}  \mathcal{L} (F)$, because they use $\frac{\partial{F_{\mu\nu}}}{\partial{F_{\rho\sigma}}} = \delta^\mu_\rho \delta^\nu_\sigma$ rather than $\frac{\partial{F_{\mu\nu}}}{\partial{F_{\rho\sigma}}} = \delta^\mu_\rho \delta^\nu_\sigma - \delta^\mu_\sigma \delta^\nu_\rho$. We will avoid this trick and differentiate by independent components of the antisymmetric tensors.} 
    \begin{equation}\label{eq:ConstitutiveEq}
        \wave{G}^{\mu\nu} \equiv \frac{\partial}{\partial F_{\mu\nu}}  \mathcal{L} (F)
    \end{equation} and the $\sim$-operation on antisymmetric tensors $X$
    \begin{equation}
        \wave{X}^{\mu_1\cdots\mu_{d-p}} \equiv \frac{1}{p!} \varepsilon^{\mu_1\cdots\mu_{d-p}\nu_1\cdots\nu_p} X_{\nu_1\cdots\nu_p}, \qquad \wave{X}_{\mu_1\cdots\mu_{d-p}} \equiv \frac{1}{p!} \varepsilon_{\mu_1\cdots\mu_{d-p}\nu_1\cdots\nu_p} X^{\nu_1\cdots\nu_p},
    \end{equation}
    which plays a central role in the Carrollian treatment of generalized duality. These two definitions depend only on the Levi-Civita tensors and apply to  the Lorentzian case as well. Different from the Hodge-$*$ operator which maps a $p$-form to an $(d-p)$-form, the $\sim$-operator maps differential forms to antisymmetric polyvectors and vice versa, and this operation does not lose any information because $\wave{\wave{X}} = (-1)^s X$.\par

    We say ``pure electrodynamics" when the Lagrangian of the electromagnetic theory $\mathcal{L}(F)$  depends on $F_{\mu\nu}$ only. The equations of motion obtained from the Lagrangian $\mathcal{L}(F)$ and the Bianchi identity are then
    \begin{equation}
        \partial_\mu \wave{G}^{\mu\nu}=0,\qquad \partial_\mu \wave{F}^{\mu\nu}=0.
    \end{equation}
    It is not difficult to observe that $F$ and $G$ are on equal footing. The most general form of infinitesimal transformations of generalized duality symmetry for pure electrodynamics is usually written  as
    \begin{equation}\label{eq:InfiGeneralizedDual}
        \delta \col{F}{G} = \begin{pmatrix}a&b\\c&d\end{pmatrix} \col{F}{G}.
    \end{equation} 
    The convention is unambiguous for relativistic theory, but to discuss Carrollian theories we should strictly distinguish the positions of indices. By using the $\sim$-operator we can define $G_{\mu\nu} \equiv (-1)^s \wave{\wave{G}^{\mu\nu}}$ from the dual field strength $\wave{G}^{\mu\nu}$. Then \eqref{eq:InfiGeneralizedDual} implies the following infinitesimal duality transformations of  $(F,G)$ and $(\wave{F},\wave{G})$:
    \begin{equation}\label{eq:InfiGeneralizedDual-indexed}
        \delta \col{F_{\mu\nu}}{G_{\mu\nu}} = \begin{pmatrix}a&b\\c&d\end{pmatrix} \col{F_{\mu\nu}}{G_{\mu\nu}}, \qquad
        \delta \col{\wave{F}^{\mu\nu}}{\wave{G}^{\mu\nu}} = \begin{pmatrix}a&b\\c&d\end{pmatrix} \col{\wave{F}^{\mu\nu}}{\wave{G}^{\mu\nu}}.
    \end{equation}  In relativistic Maxwell theory, since the Lagrangian is $\mathcal{L}(F)= -\frac{1}{4} F_{\mu\nu}F^{\mu\nu}$, the dual field strength is $\wave{G}^{\mu\nu} = -F^{\mu\nu}$ and the Hodge dual of field strength is $(*F)_{\mu\nu}= G_{\mu\nu}$. The $\text{SO}(2)$ {EM duality  is generated} by \eqref{eq:InfiGeneralizedDual-indexed} with $a=d=0,\quad b=-c$.\par
    
    The Gaillard-Zumino condition of generalized duality symmetry is just the requirement of self-consistence in the definition of duality transformation because $G$ depends on $F$ through the constitutive equation \eqref{eq:ConstitutiveEq}. It can be expressed as
    \begin{equation}
        \wave{G}^{\mu\nu} + \delta \wave{G}^{\mu\nu} = \frac{\partial \mathcal{L} (F + \delta F)}{\partial (F_{\mu\nu} + \delta F_{\mu\nu})}  .
    \end{equation} Now we have
    \begin{equation}
        \begin{aligned}
            \text{R.H.S.} &= \frac{1}{2} \frac{\partial \mathcal{L} (F + \delta F)}{\partial F_{\rho\sigma}}\cdot \frac{\partial F_{\rho\sigma}}{\partial (F_{\mu\nu} + \delta F_{\mu\nu})}\\
            &= \frac{\partial \mathcal{L} (F + \delta F)}{\partial F_{\mu\nu}} - a \frac{\partial \mathcal{L} (F)}{\partial F_{\mu\nu}} - \frac{1}{2} b \wave{G}^{\rho\sigma} \frac{\partial G_{\rho\sigma}}{\partial F_{\mu\nu}},
        \end{aligned} 
    \end{equation} and also
    \begin{equation}
        \text{L.H.S.} = \frac{\partial \mathcal{L} (F)}{\partial F_{\mu\nu}} + c \wave{F}^{\mu\nu} +d \wave{G}^{\mu\nu}.
    \end{equation} Comparing the two sides and  using 
    \begin{equation}
        \begin{aligned}
            &\frac{\partial}{\partial F_{\mu\nu}} \left(F_{\rho\sigma}\wave{F}^{\rho\sigma}\right) = 4 \wave{F}^{\mu\nu}, \qquad \frac{\partial}{\partial F_{\mu\nu}} \left(G_{\rho\sigma}\wave{G}^{\rho\sigma}\right) = 2 \wave{G}^{\rho\sigma} \frac{\partial G_{\rho\sigma}}{\partial F_{\mu\nu}},\\
            &\mathcal{L} (F+\delta F)-\mathcal{L} (F) = \frac{1}{2} \frac{\partial \mathcal{L}}{\partial F_{\mu\nu}} \delta F_{\mu\nu} = \frac{1}{2} \wave{G}^{\mu\nu}(a F_{\mu\nu} + b G_{\mu\nu}),
        \end{aligned}
    \end{equation} we get the Gaillard-Zumino consistency condition
    \begin{equation}\label{eq:GZCond-General}
        \frac{b}{4}G_{\mu\nu}\wave{G}^{\mu\nu} - \frac{c}{4}F_{\mu\nu}\wave{F}^{\mu\nu} = (a+d) \mathcal{L}(F) - \frac{a}{2}F_{\mu\nu} \wave{G}^{\mu\nu} + \text{const}.
    \end{equation} We can see that the establishment of this condition only depends on the $\sim$-operation so that it applies  not only to relativistic electromagnetism but also to Carrollian theories. Specifically, when we take $a=d=0,\quad b=-c$ and make the assertion\footnote{For the usual relativistic NED theories this assertion is achieved \cite{Gibbons:1995cv} by requiring that the Lagrangian tends to the usual Maxwell Lagrangian in the weak-field limit $F \to 0$. But since the Lagrangian of the ModMax theory is not analytic in $F$, it seems that we can not take the weak-field limit in general case. However, even in the situations that the weak-field limit is not well-defined,  we are allowed to apply this requirement if one can find any path along which $F\to0$ and the Lagrangian reduces to the Maxwell Lagrangian.  For example, in the ModMax and our Carrollian ModMax theories, we can consider the zero magnetic field configurations $F_{ij}=0$, under which the Lagrangians become analytic and have a  weak-field limit, which  leads to the conclusion that the constant in \eqref{eq:GZCond-General} is zero.}  that the constant in \eqref{eq:GZCond-General} vanishes, we arrive at the Gaillard-Zumino condition for $\text{SO}(2)$ EM duality invariance:
    \begin{equation}\label{eq:GZCond-SO(2)}
        G_{\mu\nu}\wave{G}^{\mu\nu} + F_{\mu\nu}\wave{F}^{\mu\nu} =0.
    \end{equation} It is easy to check that the Lagrangian of electric-type Carrollian Maxwell electrodynamics $\mathcal{L}=\frac{1}{2} F_{0i}^2$ does not satisfy the condition \eqref{eq:GZCond-SO(2)}.\par
    
    \subsection{Generalized duality symmetry of (electric-type) Carrollian Maxwell electrodynamics}
    
    In the previous section we have found discrete duality transformation of Carrollian Maxwell electrodynamics, it is natural to ask whether this theory possesses generalized duality. It has been shown that $\text{SO}(2)$ EM duality is not the symmetry of this theory, so we need to solve for the general form of Gaillard-Zumino condition \eqref{eq:GZCond-General}. The Lagrangian $\mathcal{L}=\frac{1}{2} F_{0i}^2$ tells that $\wave{G}^{0i} = -\wave{G}^{i0} = F_{0i}$, and other components of dual field strength vanish. Consequently, $G_{\mu\nu}\wave{G}^{\mu\nu}=0$ and $\mathcal{L}(F=0) =0, \ \wave{G}(F=0) =0$. Substituting into \eqref{eq:GZCond-General} we have
    \begin{equation}
        - \frac{c}{4}F_{\mu\nu}\wave{F}^{\mu\nu} = \frac{a-d}{2} F_{0i}^2.
    \end{equation} This condition holds if and only if $c=0$ and $a=d$. These constraints restrict the duality {transformations to the } dilation given by $b=c=0,\ a=d$ and a ``shift" $a=c=d=0$. The duality group is isomorphic to $\mathbb{R}_{>0} \times \mathbb{R}$.\par

    It can be proved that the  dilation is equivalent to the scaling symmetry when the spacetime dimension is $d=4$, both for relativistic theories and Carrollian theories. The reason is that the energy-momentum tensor of $\mathcal{L}(F)$ can be expressed as 
    \begin{equation}\label{eq:NEDEMT}
        T^\mu_{\ \nu}=-\left(\wave{G}^{\mu\rho}F_{\rho\nu}+\delta^{\mu}_{\nu}\mathcal{L}\right).
    \end{equation} 
    Without virial currents, the conservation equation of the current corresponding to the scaling symmetry requires the energy-momentum tensor to be traceless. Using the fact that the Gaillard-Zumino condition of dilation $b=c=0,\ a=d$ is
    \begin{equation}\label{eq:GZCond-dilation}
        2a\mathcal{L} = \frac{a}{2} F_{\mu\nu} \wave{G}^{\mu\nu},
    \end{equation} then it is exactly equivalent to the traceless condition
    \begin{equation}
        T^\mu_{\ \mu}= \wave{G}^{\mu\nu}F_{\mu\nu}-4\mathcal{L} =0.
    \end{equation} Moreover, the traceless condition also implies the whole Carrollian conformal symmetry, as we prove in Appendix \ref{app:CarrConformal}.\par

    On the other hand, the $\mathbb{R}$-duality shift symmetry gives the transformation in electric and magnetic fields
    \begin{equation}\label{eq:ElecShiftDual}
        \mathbf{E}_e \to \mathbf{E}_e, \qquad \mathbf{B}_e \to \mathbf{B}_e + \xi \mathbf{E}_e.
    \end{equation} The electric-type Carrollian Maxwell’s equations are invariant under these duality shift symmetries, {and the duality transformation in \eqref{eq:CarrEMDual-new} is the transformation \eqref{eq:ElecShiftDual} with $\xi=1$.}\par

    \subsection{Carrollian ModMax theories}
    
    The ModMax theories is a family of conformal $\text{SO}(2)$ EM duality invariant NED theories. They were  first discovered in \cite{Bandos:2020jsw}, and were rederived by using a much simpler method in \cite{Kosyakov:2020wxv}. In this subsection, we are going to construct Carrollian ModMax theories, which are conformal invariant and $\text{SO}(2)$ EM duality invariant.

    From now on we use $F^{\mu\nu}$ for $F^{\mu\nu} \equiv (m^{\mu\rho}\gamma^{\nu\sigma}+m^{\nu\sigma}\gamma^{\mu\rho})F_{\rho\sigma}$, and this notation will not cause any confusion because the $\sim$-operation of $F^{\mu\nu}$ satisfies $\wave{F}_{\mu\nu} = \wave{F}^{\rho\sigma} h_{\rho\mu} h_{\sigma\nu}$. Moreover, we introduce  the Carrollian invariant quantities as the building blocks of the Lagrangian,
    \begin{equation}\label{eq:CarrSP}
        \begin{aligned}
            &\mathcal{S} \equiv \frac{1}{2} F_{\mu\nu}F^{\mu\nu}=  m^{\mu\rho}\gamma^{\nu\sigma}F_{\mu\nu}F_{\rho\sigma},& 
            &\mathcal{P} \equiv \frac{1}{2} F_{\mu\nu}\wave{F}^{\mu\nu},
        \end{aligned}
        \end{equation}
        and introduce the notations
        \begin{equation}\label{eq:CarrSP2}
        \begin{aligned}
            &\mathcal{L}_\mathcal{S} \equiv \frac{\partial \mathcal{L}(\mathcal{S},\mathcal{P})}{\partial \mathcal{S}}, & &\mathcal{L}_\mathcal{P} \equiv \frac{\partial \mathcal{L}(\mathcal{S},\mathcal{P})}{\partial \mathcal{P}}.
        \end{aligned}
    \end{equation} 
    Notice that the Carrollian $\mathcal{P}$ is opposite to the Lorentz $\mathcal{P}$ due to our signature choice. For relativistic pure electrodynamics, all of the Lorentz invariant Lagrangians can be expressed as a function of relativistic $\mathcal{S},\ \mathcal{P}$ defined in \eqref{eq:ModMaxLag}, which correspond to the Maxwell Lagrangian and the $\theta$-term respectively. When it comes to the Carrollian symmetry, finding all Carrollian invariant quantities becomes more complicated because of the existence of magnetic-type quantities. However, if we exclude them by restricting ourselves to pure electrodynamics and focus on the electric-type theory, the Carrollian invariant NED Lagrangians $\mathcal{L}(\mathcal{S},\mathcal{P})$ are assumed to be functions of Carrollian $\mathcal{S},\ \mathcal{P}$. Here $\mathcal{S}$ is proportional to the Lagrangian $\mathcal{L}=\frac{1}{2} \mathcal{S}$ of electric-type Carrollian Maxwell electrodynamics, which can be checked to be locally boost invariant, and the topological term $\mathcal{P}$ is the same as in the relativistic case.\par

    By direct computation, we find  the dual field strength 
    \begin{equation}
        \wave{G}^{\mu\nu}= 2\left( \mathcal{L}_\mathcal{S} F^{\mu\nu} + \mathcal{L}_\mathcal{P} \wave{F}^{\mu\nu} \right),
    \end{equation}
    and the Gaillard-Zumino conditions of  dilation (or equivalently the scaling symmetry) and $\text{SO}(2)$ EM duality
    \begin{equation}\label{eq:ModMaxEq}
        \left\{\begin{aligned}
            &F_{\mu\nu} \wave{G}^{\mu\nu} = 4 \mathcal{L},\\
            &F_{\mu\nu}\wave{F}^{\mu\nu} + G_{\mu\nu}\wave{G}^{\mu\nu} =0,
        \end{aligned}\right.
    \end{equation}
    which  look like the equations found in \cite{Kosyakov:2020wxv}. However, recalling that in Carrollian theory $F^{\mu\nu} \wave{F}_{\mu\nu} =0$, which is not equal to $F_{\mu\nu}\wave{F}^{\mu\nu}$,  we find that the equations \eqref{eq:ModMaxEq} can be expressed in terms of $\mathcal{S},\ \mathcal{P}$
    \begin{equation}\label{eq:ModMaxEq-expand}
        \left\{\begin{aligned}
            &\mathcal{S}\mathcal{L}_\mathcal{S} + \mathcal{P}\mathcal{L}_\mathcal{P} = \mathcal{L},\\
            &4\mathcal{L}_\mathcal{P}(2\mathcal{S}\mathcal{L}_\mathcal{S} + \mathcal{P}\mathcal{L}_\mathcal{P}) + \mathcal{P}=0.
        \end{aligned}\right.
    \end{equation} 
     If we impose the reality condition on the Lagrangian\footnote{The reality condition restricts the Lagrangian to be purely real or imaginary, and it causes the $\mp$-sign to appear in the Lagrangian \eqref{eq:CarrModMaxLag}. Strictly speaking, the Lagrangian with the plus sign is a solution to \eqref{eq:ModMaxEq-expand} only if an extra overall $i$ factor is included. The Lagrangians with two signs can be converted into each other through $\gamma \to \gamma + i \frac{\pi}{2}$. At the same time, the Wick rotation $x^0 \to i x^0$ also exchanges this sign, and the overall imaginary unit factor can be interpreted as the consequence of Wick rotation.}, the only solution to this partial differential equations is
    \begin{equation}\label{eq:CarrModMaxLag}
        \mathcal{L}_{\gamma}(\mathcal{S},\mathcal{P}) = \frac{1}{4} \left(e^{\gamma}\mathcal{S} \mp e^{-\gamma}\frac{\mathcal{P}^2}{\mathcal{S}}\right),
    \end{equation} 
    where $\gamma$ is a free real parameter, as in ModMax. This one-parameter family of theories provides the only Carrollian conformal and $\text{SO}(2)$ EM duality invariant NEDs.  Obviously,  the Lagrangians \eqref{eq:CarrModMaxLag} cannot be realized as the ultra-relativistic limit of Lorentz ModMax Lagrangian
    \begin{equation}\label{eq:ModMaxLag-2}
        \mathcal{L}_\gamma (\mathcal{S},\mathcal{P}) = -\frac{1}{2} \cosh{\gamma} \mathcal{S} + \frac{1}{2} \sinh{\gamma} \sqrt{\mathcal{S}^2+\mathcal{P}^2}.
    \end{equation} \par
    
    Up to a renormalization,  the Carrollian ModMax family \eqref{eq:CarrModMaxLag} reaches two different limits as $\gamma \to \pm \infty$
    \begin{equation}
        \mathcal{L}_{+\infty}(\mathcal{S},\mathcal{P}) \sim \mathcal{S}, \qquad \mathcal{L}_{-\infty}(\mathcal{S},\mathcal{P}) \sim \frac{\mathcal{P}^2}{\mathcal{S}}.
    \end{equation} The $\mathcal{L}_{+\infty}$ limit is just the Lagrangian of electric-type Carrollian Maxwell electrodynamics, while $\mathcal{L}_{-\infty}$ is a novel Carrollian NED. Using the generalized Gaillard-Zumino condition, the $\mathcal{L}_{-\infty}$ shares the same duality symmetries as the electric-type Carrollian Maxwell electrodynamics, whose duality group is $\mathbb{R}_{>0} \times \mathbb{R}$. The duality group of Carrollian ModMax theories is $\mathbb{R}_{>0} \times \text{SO}(2)$, which is similar to the one of relativistic ModMax and Maxwell theories. Again, using the Gaillard-Zumino condition of $c=0,\ a=d$ one can check that $\mathcal{L}_{-\infty}$ is the only (pure) Carrollian NED that shares the same $(\mathbb{R}_{>0} \times \mathbb{R})$-duality symmetry with electric-type Carrollian Maxwell theory.\par

\section{Carrollian $\sqrt{T\bar{T}}$ deformation flows}\label{sec:TTbar}

    Initiated by \cite{Smirnov:2016lqw, Cavaglia:2016oda}, the $T\bar{T}$-like deformations have raised broad interest in various areas of theoretical physics because they provide important examples of solvable deformations that preserve integrability and duality symmetry. Most generally, as {shown in \cite{Conti:2022egv,Hou:2022csf}, we can consider the $T\bar{T}$-like deformation in $d$-dimensional spacetime, which is generated by an irrelevant operator},
    \begin{equation}\label{eq:TTbarOp}
        O^\lambda_{T^2} = \frac{1}{d}\left(T^\mu_{\ \nu} T^\nu_{\ \mu} - r T^\mu_{\ \mu} T^\nu_{\ \nu}\right), \hspace{3ex}r \in \mathbb{R},
    \end{equation} 
    {The name ``$T\bar{T}$ deformation" usually refers to the case $(d,r)=(2,1)$\cite{Smirnov:2016lqw, Cavaglia:2016oda} or $(d,r)=(4,\frac{1}{2})$\cite{Conti:2018jho}. When considering $T\bar{T}$-like deformation the parameter $r$ can be chosen differently for different purposes \cite{Conti:2022egv,Ondo:2022zgf,Tsolakidis:2024wut}, but it does not matter if the energy-momentum tensor is traceless. The $\sqrt{T\bar{T}}$ deformation is a deformation generated by the classically marginal operator $O^\lambda_{\sqrt{T^2}}$ \cite{Rodriguez:2021tcz, Conti:2022egv, Ferko:2022cix} with $(d,r)=(d,\frac{1}{d})$, which is the square root of \eqref{eq:TTbarOp},}
    \begin{equation}\label{eq:RootTTbarOp}
        O^\gamma_{\sqrt{T^2}} = \sqrt{\frac{1}{d}\left(T^\mu_{\ \nu} T^\nu_{\ \mu} - r T^\mu_{\ \mu} T^\nu_{\ \nu}\right)}.
    \end{equation} 
    Note that in $2D$ the square of the $\sqrt{T\bar{T}}$ operator is $T\bar{T}$-like but does not equate to the $T\bar{T}$-operator due to historical reason, and this sometimes {causes} confusion.\par
    
    As we have mentioned, relativistic ModMax theories are closely related to the Maxwell theory through the $\sqrt{T\bar{T}}$ flow \cite{Babaei-Aghbolagh:2022uij,Ferko:2022iru}. Recently, a generic geometric formalism of $\sqrt{T\bar{T}}$ deformation in arbitrary dimensions was obtained in \cite{Babaei-Aghbolagh:2024hti}. In this section, we will investigate this relationship in the context of the Carrollian ModMax theories. We find that under the $\sqrt{T\bar{T}}$ flow, the Carrollian ModMax theories reach two fixed points, one of which is the electric-type Carrollian Maxwell theory, and the flow is non-invertible at these endpoint theories. This makes the story quite different from the relativistic $\sqrt{T\bar{T}}$ flow of the ModMax theories {in which } the $\sqrt{T\bar{T}}$ flow can generate the ModMax theories from the Maxwell theory. The Carrollian ModMax theories provide a concrete and novel example of $\sqrt{T\bar{T}}$ deformation of non-Lorentzian theories. More works about non-Lorentzian $T\Bar{T}$-like theories can be found in \cite{Cardy:2018jho,Jiang:2020nnb, Hansen:2020hrs, Ceschin:2020jto, Chen:2020jdi, Esper:2021hfq}.\par
    
    Furthermore, inspired by $4D$ ModMax theories, a few research groups proposed independently in  \cite{Ferko:2022cix,Conti:2022egv,Babaei-Aghbolagh:2022leo} that a one-parameter family of $2D$ ModMax-like theories of  multiple free scalars could be generated by  the $2D$ $\sqrt{T\bar{T}}$ deformation flow from a free multi-scalar theory. We will show that the $2D$-dimensional reduction of $4D$ Carrollian ModMax theories and its generalization lead to a family of modified $2D$ multiple BMS-scalar theories\cite{Hao:2021urq}, which are also closed under the $\sqrt{T\bar{T}}$ flow.\par

\subsection{Carrollian ModMax under generalized $\sqrt{T\bar{T}}$ deformation}

    The relativistic ModMax theories \eqref{eq:ModMaxLag-2} are known \cite{Babaei-Aghbolagh:2022uij} to be generated from the Maxwell theory by the $\sqrt{T\bar{T}}$ flow perturbatively in $d=4$, in the sense that 
    \begin{equation}\label{eq:ModMaxFlow}
        \frac{\partial \mathcal{L}^{\text{ModMax}}_\gamma}{\partial \gamma} = O^\gamma_{\sqrt{T^2}}, \qquad \mathcal{L}_\gamma^{\text{ModMax}} = \mathcal{L}^{\text{Maxwell}} + \int O^\gamma_{\sqrt{T^2}} d\gamma.
    \end{equation} To convince this, we just need to substitute the ModMax Lagrangians \eqref{eq:ModMaxLag} and the expression of $\sqrt{T\bar{T}}$ operator \eqref{eq:RootTTbarOp} into the above equation and verify that it is a solution. Now we look at the Carrollian ModMax theories $\mathcal{L}^{\text{CMM}}_{\gamma}$ in \eqref{eq:CarrModMaxLag}:
    \begin{equation}\label{eq:CarrModMaxLag-2}
        \mathcal{L}^{\text{CMM}}_{\gamma}(\mathcal{S},\mathcal{P}) = \frac{1}{4} \left(e^{\gamma}\mathcal{S} \mp e^{-\gamma}\frac{\mathcal{P}^2}{\mathcal{S}}\right).
    \end{equation}
    The energy-momentum tensor of a general relativistic nonlinear electrodynamics is \cite{Kosyakov:2007qc}
    \begin{equation}\label{eq:EMT-general}
        T^\mu_{\ \nu}=-\left(\wave{G}^{\mu\rho}F_{\rho\nu}+\delta^{\mu}_{\nu}\mathcal{L}\right).
    \end{equation} This expression also works for Carrollian nonlinear electrodynamics, and we prove it in Appendix \ref{appsec:EMT-CarrNED}. We use ${(T_0)}^\mu_{\ \nu}$ as the energy-momentum tensor of the electric-type Carrollian Maxwell theory $\mathcal{L}^{\text{eCMax}}=\frac{1}{2}\mathcal{S}$,
    \begin{equation}\label{eq:EMT-eCMax}
        {(T_0)}^\mu_{\ \nu}=-\left(F^{\mu\rho}F_{\rho\nu}+\frac{1}{4}F^{\rho\sigma}F_{\rho\sigma}\delta^{\mu}_{\nu}\right).
    \end{equation} The energy-momentum tensors of the Carrollian ModMax theories are
    \begin{equation}\label{eq:EMT-CMM}
        {(T_\gamma)}^\mu_{\ \nu}= \frac{1}{2}{(T_0)}^\mu_{\ \nu}\left(e^{\gamma} \pm e^{-\gamma}\frac{\mathcal{P}^2}{\mathcal{S}^2}\right).
    \end{equation} Then the $T\bar{T}$ operators of electric-type Carrollian Maxwell theory and Carrollian ModMax theories are {respectively}
    \begin{equation}\label{eq:T2Operator}
        \begin{aligned}
            O_{T_0^2} &= \frac{1}{4}\mathcal{S}^2, \\
            O^\gamma_{T^2} &= \frac{1}{16}\mathcal{S}^2 \left(e^{\gamma} \pm e^{-\gamma}\frac{\mathcal{P}^2}{\mathcal{S}^2}\right)^2 = \left( \frac{e^{\gamma}\mathcal{S}^2 \pm e^{-\gamma}\mathcal{P}^2}{4\mathcal{S}}\right)^2.
        \end{aligned}
    \end{equation} {Therefore,} we have
    \begin{equation}\label{eq:CMMFlow}
         \frac{\partial \mathcal{L}_\gamma^{\text{CMM}}}{\partial \gamma} = \frac{1}{4} \left( e^{\gamma}\mathcal{S} \pm e^{-\gamma}\frac{\mathcal{P}^2}{\mathcal{S}} \right)= O^\gamma_{\sqrt{T^2}}.
    \end{equation}\par
    
    We observe that the family of Carrollian ModMax theories \eqref{eq:CarrModMaxLag-2} deform among themselves under $\sqrt{T\bar{T}}$ flow. As the parameter $\gamma$ tends to infinity, the endpoints of this flow, we find the Lagrangian of electric-type Carrollian Maxwell theory $\mathcal{L}\sim \mathcal{S}$ and the other $\mathbb{R}$-duality shift invariant Carrollian conformal NED $\mathcal{L}\sim \frac{\mathcal{P}^2}{\mathcal{S}}$. However, we cannot deform the endpoint theories via the $\sqrt{T\bar{T}}$ flow to get the Carrollian ModMax theories, because the flow is non-invertible at these two points. The endpoint theories are the fixed points of the $\sqrt{T\bar{T}}$ flow, because
    \begin{equation}\label{eq:EndPointFlow}
        \begin{aligned}
        O^\gamma_{\sqrt{T^2}} &= e^{\gamma}\frac{\mathcal{S}}{2} &\text{ for } &\mathcal{L} = e^{\gamma} \frac{\mathcal{S}}{2},\\
        O^\gamma_{\sqrt{T^2}} &= e^{-\gamma}\frac{\mathcal{P}^2}{2\mathcal{S}} &\text{ for } &\mathcal{L} = -e^{-\gamma}\frac{\mathcal{P}^2}{2\mathcal{S}}.
        \end{aligned}
    \end{equation} {In the relativistic case, it was discovered in \cite{Ferko:2022iru} and was later proved in \cite{Ferko:2023wyi} in detail that the $\sqrt{T\bar{T}}$ flow preserves the duality symmetry. In the Carrollian case, the duality symmetry is still preserved under the $\sqrt{T\bar{T}}$ flow except for the non-invertible endpoints. So we propose but leave the proof to future work that even for the non-Lorentzian theories, the $\sqrt{T\bar{T}}$ flow would preserve the duality symmetry although it is invertible. } The relationship between the endpoint theories and Carrollian ModMax theories is illustrated in the following diagram, and the red points contain an infinite renormalization factor:
    \begin{equation}\label{eq:CMMFlowDiagram}
        \begin{tikzcd}[row sep=tiny]
            \mathcal{L} \sim - \frac{\mathcal{P}^2}{2\mathcal{S}}&&&&\mathcal{L} \sim \frac{\mathcal{S}}{2}\\
            \textcolor{red}{\bullet} \ar[d,phantom,"-\infty"  description] \ar[rrrr,leftrightarrow,"\text{Carrollian ModMax}","\sqrt{T\bar{T}}-\gamma\text{ flow}"swap] &&&& \textcolor{red}{\bullet} \ar[d,phantom,"+\infty"  description] \\
            \  &&&& \ \\
        \end{tikzcd}.
    \end{equation}\par

\subsection{$\sqrt{T\bar{T}}$ deformation of $2D$ Carrollian modified scalar theories} \label{subsec:2dReduction}

    In the last subsection, we discuss the behavior of $4D$ Carrollian ModMax theories under the $\sqrt{T\bar{T}}$ flow, now we show that the story can be generalized to $2D$, and we construct a Carrollian ModMax-like extension of $2D$ BMS free scalar theory. To be more specific, we consider $2D$ scalar field theories, where the scalar has internal degrees of freedom and can be taken as $N$ copies of bosonic scalar fields, or in other words we consider $2D$ $\text{O}(N)$ free scalar theory. We reproduce part of the relativistic picture that there is a link between the $2D$ ModMax-like modified scalar theories  and the $4D$ ModMax theories \cite{Babaei-Aghbolagh:2022uij,Babaei-Aghbolagh:2022leo,Conti:2022egv}. For the Carrollian cases, we construct $2D$ $\sqrt{T\bar{T}}$-closed scalar theories from 4D Carrollian ModMax theories, by first doing a dimensional reduction to obtain $N=2$ scalar theories and then generalizing to the case that $N$ takes any positive integer value. Different from the relativistic cases, the $\sqrt{T\bar{T}}$ flow of the $2D$ Carrollian ModMax-like modified scalar theories have two fixed points as in \eqref{eq:CMMFlowDiagram}, where the flow is again non-invertible.\par

    In two dimensions, the $\sqrt{T\bar{T}}$ flow is generated by the operator \eqref{eq:RootTTbarOp} with $d=2$ and $r=\frac{1}{2}$, and the corresponding $T\bar{T}$-like operator \eqref{eq:TTbarOp} equals the opposite of the determinant of the traceless part of the $2D$ stress tensor $T^a_{\ b}$:
    \begin{equation}
    (O^\gamma_{\sqrt{T^2}})=\sqrt{\frac{1}{2}\left(T^a_{\ b} T^b_{\ a} - \frac{1}{2}T^a_{\ a} T^b_{\ b}\right)}=\sqrt{-\det(\hat{T}^a_{\ b})}, \qquad \hat{T}^a_{\ b} = {T}^a_{\ b}- \frac{1}{2}{T}^c_{\ c}\delta^a_{\ b}.
    \end{equation}
    We begin with a short review of the relativistic $2D$ ModMax-like scalar theories. As the relativistic ModMax theories are generated from the Maxwell theory by the $\sqrt{T\bar{T}}$ flow perturbatively, the relativistic $2D$ ModMax-like scalar theories are generated by the $2D$ $\sqrt{T\bar{T}}$ flow from a free $N$ bosonic scalar theory, which could be taken as the seed theory. The Lagrangian of the free $N$-scalar theory is given by
    \begin{equation}
        \mathcal{L}_0^{N,\text{L}}=-\frac{1}{2}\sum_{i=1}^{N}\partial_a\phi^i\partial^a\phi^i.
    \end{equation} From now on $i$ is always summed over the internal indices. The stress tensor of this theory is obtained as
    \begin{equation}\label{eq:defT}
        T^a_{\ b}=\frac{\partial\mathcal{L}}{\partial \partial_a\phi^i} \partial_b\phi^i-\delta^a_{\ b}\mathcal{L}.
    \end{equation} Under the $\sqrt{T\bar{T}}$ deformation, one can get the deformed Lagrangian:
    \begin{equation}\label{eq:LorentzNscalar}
        \mathcal{L}_\gamma^{N,\text{L}}=-\frac{\cosh{\gamma}}{2}\partial_a\phi^i\partial^a\phi^i+ \frac{\sinh{\gamma}}{2}\sqrt{2 \partial_a\phi^i\partial^b\phi^i\partial_b\phi^j\partial^a\phi^j-(\partial_a\phi^i\partial^a\phi^i)^2}.
    \end{equation}
    The deformed Lagrangian reduces to the free massless $N$-scalar theory at $\gamma=0$ after a rescaling. When $N=1$, the Lagrangian $\mathcal{L}_\gamma^{N=1,\text{L}}=e^{\gamma}\mathcal{L}_0^{N=1,\text{L}}$ is just a rescaling of the free theory Lagrangian, so the deformation is trivial. When $N>1$, due to the fact that $$\partial_a\phi^i\partial^b\phi^i\partial_b\phi^j\partial^a\phi^j \neq (\partial_a\phi^i\partial^a\phi^i)^2,$$ 
    the Lagrangian can not be rescaled to $\mathcal{L}_0^{N,\text{L}}$, so the deformation is nontrivial.

    The deformed Lagrangian \eqref{eq:LorentzNscalar} can be understood from the perspective of dimensional reduction from 4D ModMax. Recalling that the Lagrangian of Lorentzian ModMax theories is given by:
    \begin{equation}\label{eq:ModMaxLag-3}
        \mathcal{L}_\gamma (\mathcal{S},\mathcal{P}) = -\frac{1}{2} \cosh{\gamma} \mathcal{S} + \frac{1}{2} \sinh{\gamma} \sqrt{\mathcal{S}^2+\mathcal{P}^2}; \qquad \mathcal{S}\equiv \frac{1}{2} F_{\mu\nu}F^{\mu\nu},\mathcal{P}\equiv \frac{1}{2} F_{\mu\nu}(*F)^{\mu\nu}.
    \end{equation} Inspired by the connection between particular solutions relating to scattering of plane waves of the $4D$ Born-Infeld theory and the solutions of the $2D$ Nambu-Goto theory in static gauge with $(N+2)$-dimensional target space for $N=2$ \cite{Conti:2022egv}, one can perform a dimensional reduction of the {Lorentzian $4D$} ModMax theories to two dimensions. This is achieved by identifying $\phi^1=A_2,\ \phi^2=A_3$ and specifying the following field configurations,
    \begin{equation}\label{eq:ReductionConf}
    \begin{aligned}
            &A_\mu = A_\mu(x^0,x^1),\\
            &F_{01} = \partial_0 A_1 - \partial_1 A_0 = 0.
    \end{aligned}
    \end{equation} Then $F_{\mu\nu}$ has only four independent non-vanishing components:
    \begin{equation}
        F_{02}=\partial_0\phi^1,\quad F_{03}=\partial_0\phi^2,\quad F_{12}=\partial_1\phi^1,\quad F_{13}=\partial_1\phi^2,
    \end{equation} 
    and the Lorentzian invariants $\mathcal{S}$ and $\mathcal{P}$ reduce to
    \begin{equation}
    \begin{aligned}
        &\mathcal{S}=\frac{1}{2} F_{\mu\nu}F^{\mu\nu}=\sum_{i=1}^{2}\partial_a\phi^i\partial^a\phi^i,\\
        &\mathcal{P}= \frac{1}{2} F_{\mu\nu}(*F)^{\mu\nu}=-2(\partial_1\phi^1\partial_0\phi^2-\partial_0\phi^1\partial_1\phi^2), \qquad (\varepsilon^{0123}=-1).
    \end{aligned}
    \end{equation} Plugging them into \eqref{eq:ModMaxLag-3}, the reduced theories have the Lagrangian \eqref{eq:LorentzNscalar} with $N=2$.  {One can further generalize to $N$-scalar case with the Lagrangian \eqref{eq:LorentzNscalar}.} One may also introduce the following notion to write this Lagrangian more compactly
    \begin{equation}
    \begin{aligned}
        &(\Phi_N)_{ab} \equiv \sum_{i=1}^N \partial_a\phi^i\partial_b\phi^i, \qquad (a,b \in \{0,1\}),\\
        &\mathcal{L}_\gamma^{N,\text{L}}=-\frac{\cosh{\gamma}}{2}(\Phi_N)_a^{\ a} + \frac{\sinh{\gamma}}{2} \sqrt{2(\Phi_N)_{ab}(\Phi_N)^{ab}-\left((\Phi_N)_a^{\ a}\right)^2},\\
    \end{aligned}
    \end{equation} 
    It can be directly checked that {$\mathcal{L}_\gamma^{N,\text{L}}$ satisfies the $\sqrt{T\bar{T}}$ flow equation}
    \begin{equation}
    \begin{aligned}
        &\frac{\partial \mathcal{L}_{\gamma}^{N,\text{L}}}{\partial \gamma}=(O^\gamma_{\sqrt{T^2}})^{N,\text{L}}.\\
    \end{aligned}
    \end{equation}\par

    {In analogy with the Lorentzian dimensional reduction, we can finally explore the  $\sqrt{T\bar{T}}$ deformation of Carrollian scalar theories from $4D$ seed theories.} The $4D$ Carrollian ModMax Lagrangian is given by \eqref{eq:CarrModMaxLag}:
    \begin{equation}\label{eq:Carrdimreduce}
        \mathcal{L}_{\gamma}(\mathcal{S},\mathcal{P}) = \frac{1}{4} \left(e^{\gamma}\mathcal{S} \mp e^{-\gamma}\frac{\mathcal{P}^2}{\mathcal{S}}\right); \qquad \mathcal{S} \equiv \frac{1}{2} F_{\mu\nu}F^{\mu\nu}=  m^{\mu\rho}\gamma^{\nu\sigma}F_{\mu\nu}F_{\rho\sigma}, ~~\mathcal{P} \equiv \frac{1}{2} F_{\mu\nu}\wave{F}^{\mu\nu}.
    \end{equation} According to \eqref{eq:ReductionConf} we have 
    \begin{equation}
    \mathcal{S}=\sum_{i=1}^{2}\partial_{0}\phi^i\partial_{0}\phi^i,\hspace{3ex}
        \mathcal{P}=2(\partial_1\phi^1\partial_0\phi^2-\partial_0\phi^1\partial_1\phi^2), \qquad (\varepsilon^{0123}=+1).
    \end{equation}
    Plugging the above relations into \eqref{eq:CarrModMaxLag}, we get:
    \begin{equation}\label{eq:2dCMM}
    \begin{aligned}
        \mathcal{L}_{\gamma}^{N=2,\text{C}}&=
            \frac{1}{4} e^{\gamma}\partial_{0}\phi^i\partial_{0}\phi^i \mp e^{-\gamma}\frac{(\partial_1\phi^1\partial_0\phi^2-\partial_0\phi^1\partial_1\phi^2)^2}{\partial_{0}\phi^i\partial_{0}\phi^i}\\
            &=\frac{1}{4} e^{\gamma}(\Phi_2)_{00} \mp e^{-\gamma}\frac{\det(\Phi_2)}{(\Phi_2)_{00}}.\\
    \end{aligned}
    \end{equation}
    It is easy to check that
    \begin{equation}
        \frac{\partial \mathcal{L}_{\gamma}^{N=2,\text{C}}}{\partial \gamma}=(O^\gamma_{\sqrt{T^2}})^{N=2,\text{C}}.
    \end{equation} We can then generalize the result to obtain the Carrollian $\sqrt{T\bar{T}}$-deformed Lagrangian of $N$-scalars,
    \begin{equation}\label{eq:CarrollianL}
        \mathcal{L}_{\gamma}^{N,C}=
        \frac{1}{4} e^{\gamma}(\Phi_N)_{00} \mp e^{-\gamma}\frac{\det(\Phi_N)}{(\Phi_N)_{00}}.
    \end{equation}
     The $N=1$ theory is trivial because $\det \Phi_1 = 0$. 
     For $N\geq 2$, this Lagrangian is nontrivial because as we can see from \eqref{eq:2dCMM} that $\det \Phi_2 = (\partial_1\phi^1\partial_0\phi^2-\partial_0\phi^1\partial_1\phi^2)^2 \neq 0$, and it satisfies
    \begin{equation}
        \frac{\partial \mathcal{L}_{\gamma}^{N,C}}{\partial \gamma}=(O^\gamma_{\sqrt{T^2}})^{N,C}.
    \end{equation}\par

    However, just as in the discussion of the $\sqrt{T\bar{T}}$ flow of the $4d$ Carrollian ModMax theories in the last subsection, we cannot generate the whole flow from the Carrollian Maxwell theory as seed theory because it lies on a fixed point of the $\sqrt{T\bar{T}}$ flow. After dimensional reduction from $4D$ Carrollian Maxwell theory, we find the $2D$ BMS $N$-free scalar theory, which is invariant under $\sqrt{T\bar{T}}$ flow \cite{He:2024yzx}, 
    \begin{equation}
        \mathcal{L}_0^{N,\text{C}}=\frac{1}{2}\sum_{i=1}^{N}\partial_{0}\phi^i\partial_{0}\phi^i,\hspace{3ex}
        \mathcal{L}_\gamma^{N,\text{C}}=e^{\gamma}\mathcal{L}_0^{N,\text{C}}.
    \end{equation} 
    This theory cannot be taken as a seed theory to generate other theories on the flow.
    
    In short, the Lorentzian $\sqrt{T\bar{T}}$ flow  and the Carrollian $\sqrt{T\bar{T}}$ flow are again very different in $2D$. For the  Lorentzian case, we can get the deformed Lagrangian by using the free scalar theory as the seed theory. However, for the Carrollian case, the BMS free scalar theory $\mathcal{L}^{N,C} = \frac{1}{2}{\left(\Phi_N\right)_{00}}$ and the scalar theory with $\mathcal{L}^{N,C} = \frac{\det\left(\Phi_N\right)}{\left(\Phi_N\right)_{00}}$ are fixed points at $\gamma=\pm \infty$ of the $\sqrt{T\bar{T}}$ deformation flow. We can  generate the Carrollian $\sqrt{T\bar{T}}$-deformed Lagrangian from any theory living on the flow except the endpoint theories. We show this $\sqrt{T\bar{T}}$ flow in the following diagram:
    \begin{equation}\label{eq:ScalarFlowDiagram}
        \begin{tikzcd}[row sep=tiny]
            \mathcal{L} \sim \frac{\det\left(\Phi_N\right)}{\left(\Phi_N\right)_{00}} &&&&&&\mathcal{L} \sim (\Phi_N)_{00} \\
            \textcolor{red}{\bullet} \ar[d,phantom,"-\infty"  description] \ar[rrrrrr,leftrightarrow,"\text{Carrollian modified $N$ scalar theories}","\sqrt{T\bar{T}}-\gamma\text{ flow}"swap] &&&&&& \textcolor{red}{\bullet} \ar[d,phantom,"+\infty"  description] \\
            \  &&&&&& \ \\
        \end{tikzcd}.
    \end{equation}\par

\section{Discussions}\label{Discussions}

    In this paper, we constructed a one-parameter family of Carrollian ModMax theories and studied their properties under  the Carrollian electromagnetic duality transformations and the $\sqrt{T\bar{T}}$ flow.  Firstly, We 
    noticed that the Carrollian EM duality transformation could be defined with the help of the non-Riemannian Hodge duality. Secondly, we used the Carrollian Gaillard-Zumino consistency condition for Carrollian NEDs to show that the Carrollian Maxwell theory is invariant under dilation and  shift, which is different from the well-known transformation originating from the Carrollian limit of the EM duality. Furthermore, we showed that if we focused on pure electrodynamics, the Carrollian ModMax theories are the only Carrollian conformal and $\text{SO}(2)$ EM duality invariant NEDs. Finally we discussed the $\sqrt{T\bar{T}}$ flow of Carrollian theories. We showed that the Carrollian ModMax theories could flow to the Carrollian Maxwell theory by the $\sqrt{T\bar{T}}$ deformation. However, the Carrollian Maxwell theory is the fixed point under the $\sqrt{T\bar{T}}$ flow, and cannot lead to Carrollian ModMax theories. In other words, the $\sqrt{T\bar{T}}$ deformation is not invertible at the endpoints of the flow. This is very different from the usual picture between the Lorentzian ModMax theories and Maxwell theory, which are connected by the $\sqrt{T\bar{T}}$ flow.   Moreover, we investigated the $\sqrt{T\bar{T}}$ flow of a family of $2D$ Carrollian ModMax-like $N$-scalar models, whose construction is inspired by the dimensional reduction of the $4D$ Carrollian ModMax theories, and we found that they are driven to the $2D$ BMS free $N$-scalar model by the flow.

    There are still several important issues to discuss. First of all, although our discussion is based on Carrollian symmetry and Carrollian NEDs, we can generalize our discussion to the Galilean NEDs without any difficulty. The Galilean analog of Carrollian structure is the Newton-Cartan structure which is given by the pair $(\gamma^{\mu\nu},\tau_\mu)$. Here $\gamma^{\mu\nu}$ is also degenerate and $\tau_\mu$ is a one-form belonging to its kernel. Covariant Newton-Cartan theories should be invariant under the Galilean cousins of Carrollian local boosts - the Milne boosts parameterized by $b^{\mu}(x)$ that act on the inverse $(h_{\mu\nu},n^\mu)$ of Newton-Cartan structure as
    \begin{equation}\label{eq:MilneBoost}
        n^\mu \to n^\mu + b^\mu, \quad h_{\mu\nu} \to h_{\mu\nu} + 2 b_{(\mu}\tau_{\nu)} \qquad \text{where} \qquad \tau_\mu b^\mu=0, \quad b_\mu = h_{\mu\nu}b^\nu.
    \end{equation}
    One can similarly define the Levi-Civita tensors through the determinant of 
    \begin{equation}
        (g_\tau)_{\mu\nu} \equiv (-1)^s \tau_\mu\tau_\nu + h_{\mu\nu}, 
    \end{equation}which is invariant for a given Newton-Cartan structure, and discuss the Galilean Hodge duality and the continuous generalized duality symmetries for Galilean NEDs. The only quadratic gauge and Galilean invariant electromagnetic theory has the Lagrangian\footnote{This Lagrangian however is non-dynamical and it is not the Lagrangian of \textit{Galilean electrodynamics} (GED) \cite{Santos:2004pq,Festuccia:2016caf,Bagchi:2022twx}, which includes an extra scalar field. Different from the Carrollian case, the two Galilean limits of Maxwell’s equations both have no simple Lagrangian descriptions, instead, the GED Lagrangian is more commonly considered, which is closely related to the magnetic Galilean limit of Maxwell’s equations. Our discussion will need more adjustments to deal with the GED theory. } \cite{Chapman:2020vtn}:
    \begin{equation}\label{eq:GEM}
        \mathcal{L} = \frac{1}{2} F_{ij}F^{ij}.
    \end{equation}  We can take $F^{\mu\nu} \equiv \gamma^{\mu\rho}\gamma^{\nu\sigma}F_{\rho\sigma}$ and find the Galilean invariant quantities
    \begin{equation}
        \mathcal{S} = \frac{1}{2}F_{\mu\nu}F^{\mu\nu} = \frac{1}{2}\gamma^{\mu\rho}\gamma^{\nu\sigma}F_{\mu\nu}F_{\rho\sigma}, \qquad \mathcal{P} = \frac{1}{2} F_{\mu\nu}\wave{F}^{\mu\nu}.
    \end{equation} General Galilean NEDs without extra fields are given by $\mathcal{L}(\mathcal{S},\mathcal{P})$, and their duality symmetries are also determined by the Gaillard-Zumino condition \eqref{eq:GZCond-General}. By solving it, the generalized duality group of Galilean electromagnetic theory \eqref{eq:GEM} is again $\mathbb{R}_{>0} \times \mathbb{R}$, consisting of the { dilation and Galilean  shift.} The Lagrangian of Galilean NEDs with generalized duality group $\mathbb{R}_{>0} \times \text{SO}(2)$ is
    \begin{equation}\label{eq:GalModMaxLag}
        \mathcal{L}^{\text{GMM}}_{\gamma}(\mathcal{S},\mathcal{P}) = \frac{1}{4} \left(e^{\gamma}\mathcal{S} \mp e^{-\gamma}\frac{\mathcal{P}^2}{\mathcal{S}}\right).
    \end{equation} These theories are the only Galilean NEDs that are Galilean conformal and $\text{SO}(2)$ EM duality invariant. It can be observed that they have the same expression as Carrollian ModMax theories if we replace the Carrollian invariant quantity $\mathcal{S}$ with its Galilean version. The quantity $\mathcal{P}$ remains the same. Moreover, they are also closed under $\sqrt{T\bar{T}}$ deformation and have \eqref{eq:GEM} as one of the endpoints {under the flow}. The dimensional reduction to two dimensions as in Section \ref{subsec:2dReduction} gives us the same result for the Galilean ModMax theories, with only the {indices $0$ being replaced by  $1$.} It reflects the fact that the $2D$ Galilean conformal algebra (GCA) is isomorphic to the $2D$ Carrollian conformal algebra (CCA).
        
    Secondly, the original form of the Carrollian ModMax-like scalar theories \eqref{eq:CarrollianL} is hard to be quantized because of the appearance of the denominator. The general Carrollian ModMax-like scalar theories can be linearized classically with the help of a Langangian-multiplier field $\eta$ and a world-sheet metric $h_{ab}$ as
    \begin{equation}
    \begin{aligned}
        \mathcal{L} &\sim \frac{1}{4} e^{\gamma}  \Phi_{00} \pm 4 e^{-\gamma} \left(\eta^2 \Phi_{00} - \eta \sqrt{-\det(\Phi)}\right),\\
        &\sim \left(\frac{1}{4} e^{\gamma} \pm 4 e^{-\gamma} \eta^2\right)\Phi_{00} \mp 2 e^{-\gamma}\eta \left(-h\right)^{\frac{1}{2}} h^{ab} \Phi_{ab}. \qquad a,b\ =\ 0,1\\
    \end{aligned}
    \end{equation} It will be interesting to explore the connection between this linearized form and the Polyakov world-sheet action of the conventional string. Besides, we have discussed the $\sqrt{T\bar{T}}$ deformation of the Carrollian ModMax-like scalar theories. It would be interesting to discuss the $\sqrt{T\bar{T}}$ deformation of BMS fermionic fields \cite{Yu:2022bcp,Hao:2022xhq,Banerjee:2022ocj}.

    In the end, we want to comment on the magnetic-type Carrollian Maxwell theory. Although the magnetic-type theory is not the focus of this paper, the magnetic-type Carrollian Maxwell’s equations also acquire a duality transformation as
    \begin{equation}\label{eq:MagEMDual}
        \mathbf{E}_m \to \mathbf{E}_m + \mathbf{B}_m, \qquad \mathbf{B}_m \to \mathbf{B}_m.
    \end{equation} 
    To put them in a covariant way needs some work.  The magnetic-type Lagrangian is\footnote{Here, for simplicity, we have omitted a decoupled $\Pi_0^2$-term that would be necessary to self-consistently realize manifest off-shell Carrollian invariance on the Lagrangian. It is a consequence of the requirement that $[\delta_{B_i},\delta_{B_j}]=\delta_{[B_i,B_j]}=0$.} \cite{Chen:2023pqf}:
    \begin{equation}\label{eq:MagCarrAction}
        \begin{aligned}
            \mathcal{L}[A_i,A_0;\Pi_i] = \Pi_i (\partial_i A_0 - \partial_0 A_i) +\frac{1}{4}(\partial_i A_j - \partial_j A_i)^2.\\
        \end{aligned}
    \end{equation}
    The magnetic-type electric fields $\mathbf{E}_m$ should no longer be interpreted as part of the electromagnetic tensor $F_{\mu\nu}=2\partial_{[\mu}A_{\nu]}$, but to be identified with the multiplier fields $\Pi_i$. And there is an additional constraint $\iota_{\hat n} F =0$  to put the temporal part of $F_{\mu\nu}$ to zero, where $\iota_{\hat n}$ is the inner derivative on the differential form given by $\hat{n}$. This makes finding the covariant expression of the magnetic-type duality transformation a more subtle task. To address this concern, the correct treatment is to encode the magnetic-type electric and magnetic fields $\mathbf{E}_m,\mathbf{B}_m$ into the dual field strengths, which is defined as $\wave{G}^{\mu\nu} \equiv \frac{\partial}{\partial F_{\mu\nu}}  \mathcal{L} (F;\Pi)$. Now the magnetic-type electric and magnetic fields satisfy
    \begin{equation}\label{eq:MagEBtoF}
        \wave{G}^{\mu\nu} = \Pi_i \partial_i\wedge \partial_t+ F_{ij} \partial_i \wedge \partial_j = E_m^i \partial_i\wedge \partial_t+ \frac{1}{2} \varepsilon_{ijk}B_m^i \partial_j \wedge \partial_k.
    \end{equation} 
    Next define $G_{\mu\nu} \equiv \frac{1}{2} \varepsilon_{\mu\nu\rho\sigma} \wave{G}^{\rho\sigma}$, and one can check that using the Carrollian Hodge-$*$, the  magnetic-type Carrollian Maxwell’s equations {can be recast into the forms}
    \begin{equation}\label{eq:MagHodgeCarrMaxwell}
        d*G = 0, \qquad d G=0, \qquad (\iota_{\hat{n}} F =0).
    \end{equation}  These equations are invariant under the following duality transformations 
    \begin{equation}\label{eq:CarrEMDual-Mag}
        G \to G + *G, \qquad *G \to *G, \qquad (F \to F),
    \end{equation} 
    which gives \eqref{eq:MagEMDual}. Recalling that there is a swapping duality between the electric-type and magnetic-type Carrollian Maxwell theories, this magnetic-type discrete duality can be viewed as the combination of the swapping duality and the electric-type discrete self-duality. To discuss the generalized duality symmetry of magnetic-type theory in detail, it is necessary to study the Gaillard-Zumino condition for interacting Carrollian theories \cite{Gaillard:1981rj}, since we have not only the electromagnetic fields. Besides, it is also important to discuss the sourced Carrollian electromagnetic theories to establish the full picture of Carrollian electromagnetic duality. We leave these issues to future work.

\section*{Acknowledgments}
    We are grateful to Roberto Tateo for some helpful discussions. We would like to thank Yu-fan Zheng for discussions about Carrollian electromagnetic duality and Yunfeng Jiang, Miao He and Yi-Jun He for discussions about $\sqrt{T\bar{T}}$. We also thank Zezhou Hu, Yunsong Wei for other valuable discussions. This research is supported by {NSFC Grant  No. 11735001, 12275004.} Jue Hou is also supported by the China Postdoctoral Science Foundation under Grant Number 2024M750404 and the Jiangsu Funding Program for Excellent Postdoctoral Talent. \par
    
    \vspace{2cm}

\appendix
\renewcommand{\appendixname}{Appendix~\Alph{section}}

\section{Criterion of Carrollian conformal symmetry}\label{app:CarrConformal}

    In this appendix, we discuss the Carrollian version of the Bessel-Hagen criterion for  {conformal field theory (CFT) to show that on the flat background the scale} invariant Carrollian NED is also Carrollian conformal invariant if there exists a proper traceless energy-momentum tensor. We also exhibit how to obtain the energy-momentum tensor \eqref{eq:NEDEMT} of Carrollian NED without a non-degenerate Riemann metric.

    \subsection{Carrollian Bessel-Hagen criterion}\label{appsec:CarrBHCri}

    The Bessel-Hagen criterion \cite{Bessel1921} tells us that in relativistic field theory, the existence of a traceless symmetric energy-momentum tensor 
    \begin{equation}\label{eq:BHCriterion}
    T^\mu_{\ \mu} = 0,\qquad (T^{\mu\nu}=T^{\nu\mu})
    \end{equation} 
    implies the whole conformal symmetry. However, the energy-momentum tensor of a Carrollian theory cannot be arbitrarily raised or lowered and cannot be made symmetric. It remains to figure out whether the condition $T^\mu_{\ \mu} = 0$ implies the whole Carrollian conformal symmetry. We aim to prove a Bessel-Hagen-like criterion for Carrollian theory.\par
    
    The relativistic symmetric energy-momentum tensor $T^{\mu\nu}$ has an definition through the variation of metric $g_{\mu\nu}$ by
    \begin{equation}\label{eq:LorentzEMT}
        \delta S = \int d^n x \ \delta\left(\sqrt{||g||} \mathcal{L}(\text{matter};g) \right)= \int \sqrt{||g||}d^n x \ \frac{1}{2}T^{\mu\nu}\delta g_{\mu\nu}.
    \end{equation} 
    For infinitesimal transformation generated by a vector $\xi^\mu$, the metric changes by
    \begin{equation}
        \delta_{\hat \xi} g_{\mu\nu} = (L_{\hat \xi} {\hat g})_{\mu\nu} = 2\nabla_{(\mu}\xi_{\nu)},
    \end{equation} 
    so the action is invariant under all conformal transformations satisfying $(L_{\xi} g)_{\mu\nu} = \lambda g_{\mu\nu}$, 
    \begin{equation}
        \delta S = \int \sqrt{-g}d^n x \ T^{\mu\nu} \lambda g_{\mu\nu} = \lambda \int \sqrt{-g}d^n x \ T^\mu_{\ \mu} = 0,
    \end{equation}
    {provided that the energy-momentum tensor is traceless.} \par

    However, we need to be careful when dealing with the indices for Carrollian field theories. Without raising or lowering any indices, the energy-momentum tensor refers to such a tensor that for any Killing vector $\xi^\mu$, the action varies as
    \begin{equation}
        \delta S = \int \mathcal{E}d^n x \ T^\mu_{\ \nu}\nabla_\mu \xi^\nu.
    \end{equation} Here we use $\mathcal{E}$ for $\sqrt{||g_\tau||}$ in the Carrollian volume form \eqref{eq:CarrLCTensor-1}. The $T^\mu_{\ \nu}$ can be chosen to be the canonical energy-momentum tensor in the Noether process, but it is not symmetric.
    Another way to derive it is by relating it to the variation of the pair $(n^\mu,h_{\mu\nu})$. Some authors prefer to do the variation of the Carrollian ``metric" with regard to pair $(\tau_\mu,h_{\mu\nu})$ or $(n^\mu,\gamma^{\mu\nu})$ \cite{deBoer:2023fnj}, but we do the variation using exactly the Carrollian structure data $(n^\mu,h_{\mu\nu})$
    \begin{equation}\label{eq:VarCarrMetric}
        \delta S = \int d^n x \ \delta\left(\mathcal{E} \mathcal{L}(\text{matter};n^\mu,h_{\mu\nu}) \right)= \int \mathcal{E}d^n x \ \left(-{\mathcal{T}_n}_\mu \delta n^\mu + \frac{1}{2}{\mathcal{T}_h}^{\mu\nu}\delta h_{\mu\nu}\right).
    \end{equation}
    such that 
    \begin{equation}\label{eq:CarrEMT-metric}
        T^{\mu}_{\ \nu} \equiv -\left({\mathcal{T}_n}_\nu n^\mu + {\mathcal{T}_h}^{\mu\rho}h_{\rho\nu}\right).
    \end{equation} The ${\mathcal{T}_h}^{\mu\nu}$ is symmetric and determined up to an arbitrary term proportional to $n^\mu n^\nu$. It should be pointed out here that the variation by $n_\mu$ and $h_{\mu\nu}$ are not independent because $n^\mu h_{\mu\nu}=0$, so ${\mathcal{T}_n}_\mu$ and ${\mathcal{T}_h}^{\mu\nu}$ are not uniquely determined. Actually we have $h_{\mu\nu} \delta n^\mu + n^\mu \delta h_{\mu\nu}=0$, so the following transformation gives equivalent energy-momentum tensors for any $a^\mu$:
    \begin{equation}
        \left\{
        \begin{aligned}
            &{\mathcal{T}_n}_\mu \to {\mathcal{T}_n}_\mu + a^\mu h_{\mu\nu},\\
            &{\mathcal{T}_h}^{\mu\nu} \to {\mathcal{T}_h}^{\mu\nu} + 2n^{(\mu}a^{\nu)}.\\
        \end{aligned}
        \right.
    \end{equation}
    Then for Carrollian conformal Killing vector $\xi^\mu$, we have
    \begin{equation}\label{eq:ConfCarrKilling}
        (L_{\hat \xi} {\hat n})^\mu = - \alpha n^\mu,\qquad (L_{\hat \xi} {\hat h})_{\mu\nu} = 2 \alpha h_{\mu\nu}.
    \end{equation} And we also have
    \begin{equation}\label{eq:MetricLieDerivativeCovariant}
        (L_{\hat \xi} {\hat n})^\mu = \xi^\nu \nabla_\nu n^\mu - n^\nu \nabla_\nu \xi^\mu,\qquad (L_{\hat \xi} {\hat h})_{\mu\nu} = 2 \nabla_{(\mu}\xi_{\nu)}.
    \end{equation} Here $\nabla_\mu$ is any torsionless connection adapted to Carrollian structure $(n^\mu,h_{\mu\nu})$ with vanishing extrinsic curvature $K_{\mu\nu} = \frac{1}{2}(L_{\hat n} {\hat h})_{\mu\nu}$, 
    \begin{equation}
        \nabla_\rho n^\mu = \nabla_\rho h_{\mu\nu} = 0.
    \end{equation}
    Here $\xi_\mu$ is lowered by $h_{\mu\nu}$. Then the variation of the action is
    \begin{equation}
        \begin{aligned}
                \delta_\xi S &= \int \mathcal{E}d^n x \ \left({\mathcal{T}_n}_\nu (L_{\hat \xi} {\hat n})^\nu + \frac{1}{2}{\mathcal{T}_h}^{\mu\nu}(L_{\hat \xi} {\hat h})_{\mu\nu}\right).\\
                &= \int \mathcal{E}d^n x \ \left(\left({\mathcal{T}_n}_\nu n^\mu + {\mathcal{T}_h}^{\mu\rho}h_{\rho\nu}\right)\nabla_\mu \xi^\nu + {\mathcal{T}_n}_\nu \xi^\mu \nabla_\mu n^\nu\right)\\
                &= \int \mathcal{E}d^n x \ T^\mu_{\ \nu}\nabla_\mu \xi^\nu +0.
        \end{aligned}
    \end{equation} 
    { In this way  one can prove} that the tensor defined in \eqref{eq:CarrEMT-metric} satisfies the definition of an energy-momentum tensor. Actually, the energy-momentum tensor \eqref{eq:NEDEMT},\eqref{eq:EMT-general} can be given in this way. 

    We require the background Carrollian metric to have vanishing extrinsic curvature because only in this case the torsionless adapted connection $\nabla_\mu$ exist. The flat Carrollian metric trivially satisfies this requirement, as well as any magnetic-type Carrollian gravity solutions \cite{Hansen:2021fxi}. When the extrinsic curvature $K_{\mu\nu}$ does not vanish, we can only find torsional connections such that \eqref{eq:MetricLieDerivativeCovariant} will involve $K_{\mu\nu}$. Such torsional connections invalidate the arguments here.
    
    Now substituting \eqref{eq:ConfCarrKilling} into \eqref{eq:VarCarrMetric} we get for Carrollian conformal Killing vectors,
    \begin{equation}
        \begin{aligned}
            \delta_\xi S &= \alpha \int \mathcal{E}d^n x \ \left({\mathcal{T}_n}_\mu n^\mu + {\mathcal{T}_h}^{\mu\nu} h_{\mu\nu}\right).\\
            &= \alpha \int \mathcal{E}d^n x \ T^\mu_{\ \mu}.
        \end{aligned}
    \end{equation} In this way, we find the Bessel-Hagen-like criterion of Carrollian conformal invariance that a Carrollian field theory is Carrollian conformal invariant if the energy-momentum tensor given by variation of metric data \eqref{eq:CarrEMT-metric} is traceless $T^\mu_{\ \mu}=0$.
    
    \subsection{Energy-momentum tensor of Carrollian NED}\label{appsec:EMT-CarrNED}

    In this section, we show how to use the method in the last section to find the energy-momentum tensor of {scale} invariant Carrollian NED as \eqref{eq:NEDEMT}. We should do the variation of the Lagrangian of Carrollian NED according to the Carrollian structure. We need to use the following identities:
    \begin{equation}
        \begin{aligned}
            \delta \mathcal{E}&=\frac{1}{2}\mathcal{E} {g_\gamma}^{\mu\nu}\delta {(g_\tau)}_{\mu\nu} =  \frac{1}{2}\mathcal{E} \left( 2 n^\mu \delta \tau_\mu + \gamma^{\mu\nu} \delta h_{\mu\nu}\right)\\
            &\simeq \mathcal{E} \left( -\tau_\mu \delta n^\mu + \frac{1}{2} \gamma^{\mu\nu} \delta h_{\mu\nu} \right),\\
            \delta {\mathcal{S}}&= \delta \left(  m^{\mu\rho}\gamma^{\nu\sigma}F_{\mu\nu}F_{\rho\sigma} \right) = \frac{1}{2} \delta \left(  h_{\mu\rho}h_{\nu\sigma}\wave{F}^{\mu\nu}\wave{F}^{\rho\sigma} \right)\\
            &\simeq -2\mathcal{S}\left( -\tau_\mu \delta n^\mu +\frac{1}{2} \gamma^{\mu\nu} \delta h_{\mu\nu} \right) +  \left(h_{\nu\sigma}\wave{F}^{\mu\nu}\wave{F}^{\rho\sigma} \right) \delta h_{\mu\rho},\\
            \delta (\mathcal{E}\mathcal{P})&=\mathcal{P} \delta \mathcal{E}+ \mathcal{E} \delta \mathcal{P}=0, \qquad\text{($\mathcal{P}$ is topological)}.\\
        \end{aligned}
    \end{equation} Here $\simeq$ means equal up to the term proportional to $n^\mu n^\nu$.

    Moreover, for general $\mathcal{L}=\mathcal{L}(\mathcal{S},\mathcal{P})$, noticing that
    \begin{equation}
        \mathcal{P}\delta^\mu_{\ \nu} = -2\wave{F}^{\mu\rho}F_{\rho\nu}, 
        \qquad \mathcal{S}\delta^\mu_{\ \nu} = -F^{\mu\rho}F_{\rho\nu} - \wave{F}^{\mu\rho}\wave{F}_{\rho\nu},
    \end{equation}we have
    \begin{equation}
        \delta S = \int d^4 x \ \mathcal{L} \delta \mathcal{E} + \mathcal{E} \mathcal{L}_{\mathcal{S}} \delta \mathcal{S} - \mathcal{P} \mathcal{L}_{\mathcal{P}} \delta \mathcal{E}.
    \end{equation} 
    So {the components of the energy-momentum tensor are}
    \begin{equation}
        \begin{aligned}
            {\mathcal{T}_n}_\mu &= (2\mathcal{S}\mathcal{L}_\mathcal{S}-\mathcal{L} + \mathcal{P} \mathcal{L}_{\mathcal{P}})\tau_\mu,\\
            {\mathcal{T}_h}^{\mu\nu} &= -(2\mathcal{S}\mathcal{L}_\mathcal{S}-\mathcal{L} + \mathcal{P} \mathcal{L}_{\mathcal{P}}) \gamma^{\mu\nu} + 2\mathcal{L}_\mathcal{S} h_{\rho\sigma}\wave{F}^{\mu\rho}\wave{F}^{\nu\sigma}.\\
            -T^\mu_{\ \nu} &= -(2\mathcal{S}\mathcal{L}_\mathcal{S}-\mathcal{L}+\mathcal{P} \mathcal{L}_{\mathcal{P}})\left(\gamma^{\mu\rho}h_{\rho\nu}+n^\mu\tau_\nu\right) + 2\mathcal{L}_\mathcal{S}\wave{F}^{\mu\rho}\wave{F}_{\nu\rho} \\
            &= \mathcal{L}\delta^\mu_{\ \nu} - \left(2\mathcal{S}\mathcal{L}_\mathcal{S}+\mathcal{P} \mathcal{L}_{\mathcal{P}}\right)\delta^\mu_{\ \nu} - 2\mathcal{L}_\mathcal{S}\wave{F}^{\mu\rho}\wave{F}_{\rho\nu}\\
            & = \mathcal{L}\delta^\mu_{\ \nu} + \left( 2 \mathcal{L}_\mathcal{S} F^{\mu\rho} + 2 \mathcal{L}_\mathcal{P} \wave{F}^{\mu\rho}\right)F_{\rho\nu}\\
            &= \wave{G}^{\mu\rho}F_{\rho\nu} + \mathcal{L}\delta^\mu_{\ \nu}.
        \end{aligned}
    \end{equation}

\bibliographystyle{JHEP}
\bibliography{refs.bib}
\end{document}